\documentclass[prx,superscriptaddress,twocolumn,footinbib]{revtex4-2}
\pdfoutput=1
\usepackage[colorlinks=true,urlcolor=blue]{hyperref}
\usepackage{amsfonts}
\usepackage{graphicx}
\usepackage{placeins}
\usepackage{amsmath}
\usepackage{xcolor}
\usepackage{color}
\usepackage{bm}
\usepackage[normalem]{ulem}
\usepackage{array}
\usepackage{babel}
\usepackage{ragged2e}
\definecolor{red}{rgb}{0.75,0,0}
\definecolor{green}{rgb}{0,0.5,0}

\newcommand{\p}{{\mathfrak{p}}}
\newcommand{\q}{{\mathfrak{q}}}

\begin{document}
	
	\title{Mixed symmetries in active non-reciprocal systems}

	\author{Yuv Agarwal}
	\affiliation{Department of Theoretical Physics, University of Geneva, 1205 Geneva, Switzerland}
	
	\author{Ylann Rouzaire}
    \affiliation{Departament de F\'isica de la Materia Condensada, Universitat de Barcelona, Mart\'i i Franqu\`es 1, E08028 Barcelona, Spain}
    \affiliation{UBICS University of Barcelona Institute of Complex Systems , Mart\'i i Franqu\`es 1, E08028 Barcelona, Spain}
	
	\author{Daniel J. G. Pearce}
    \email{daniel.pearce@unige.ch}
	\affiliation{Department of Theoretical Physics, University of Geneva, 1205 Geneva, Switzerland}

 \date{\today}
 
\begin{abstract}

    Active materials are driven out of equilibrium at the microscopic scale and display diverse phenomenology; non-reciprocity is a subclass of activity in which action-reaction symmetry is also broken. 
    These models are often grouped by the order of symmetry of their components, such as polar bird flocks or hexatic cellular tissues. However, many biological examples exist in which the components appear to have two or more symmetries simultaneously. 
    To study this effect we introduce two independently controllable symmetries to the non-reciprocal XY model. This lattice model demonstrates wide ranging behavior dictated by the chosen symmetries. 
    We derive a full hydrodynamic description of our lattice model that highlights the mechanisms driving its behavior. 
    Each combination of symmetries can lead to a different hydrodynamic model with one or two simultaneous order parameters with different levels of orientational order that can appear active or not.

\end{abstract}

	\maketitle

Active matter is the study of materials in which detailed balance is intrinsically broken. This is often achieved by subunits of the material converting some internal energy source into mechanical forces, with numerous examples across length scales~\cite{shaebani2020computational,needleman2017active}. Active materials are capable of generating a wide array of behaviors that are commonly grouped by the order of symmetry
of their components, from isotropic~\cite{zhang2023pulsating, banerjee2024hydrodynamics, koehler2026flow, casagrande2026topology}, polar~\cite{vicsek1995novel,toner1995long, toner2012birth}, nematic~\cite{decamp2015orientational,giomi2015geometry,doostmohammadi2018active}, and higher orders~\cite{maitra2020chiral,giomi2022hydrodynamic,yang2026active}. Recent studies have also shown that many active materials possess mixed symmetries. For example, epithelial tissues contain both nematic and hexatic order~\cite{armengol2023epithelia, armengol2024hydrodynamics}. Self-propelled cells and cytoskeletal filaments contain mixed polar and nematic symmetries that can lead to spontaneous chirality, stabilized grain boundaries and defect lattices~\cite{beer2025Irreversibility,dinelli2026active,dumoulin2025defect}. 

An emerging subclass of active materials are non-reciprocal materials, in which action-reaction symmetry is also broken~\cite{weis2025generalizedNRmultipopulation}. A classical example of a non-reciprocal system is a predator chasing a prey, in which the mismatch in attraction and repulsion between the two animals leads to spontaneous translation~\cite{weis2025generalizedNRmultipopulation}. Such multi-species non-reciprocal models are well studied at both microscopic and hydrodynamic scales and have been shown to give rise to diverse phenomenology including spontaneous advection, synchronization, and pattern formation~\cite{dinelli2023non,weis2025generalizedNRmultipopulation,fruchart_non-reciprocal_2021}. 

Non-reciprocity in single species models often arises in asymmetric information sharing between agents for instance introduced by anisotropic sensing. This can be employed to achieve self-assembly and pattern formation~\cite{osat2023non,osat2024escaping,wu2026nonlocal}. Again taking inspiration from the animal kingdom, this is most commonly achieved by the employment of \emph{vision cones} in which agents only react to what they can ``see''. This introduces non-reciprocity as it is not necessary that an agent is able to see all the agents that can see it. This is commonly studied in flocking models~\cite{couzin2005effective,hemelrijk2011some} as well as spin lattice models~\cite{garces2025phase, rajeev2024ising,loos2023long, rouzaire2025nonreciprocal,rouzaire2026dynamics, bandini2025xy, popli2025ordering, dadhichi_nonmutual_2020, huang2024active}. 

These spin lattice models, often referred to as the non-reciprocal XY model, provide the perfect platform on which to study mixing symmetries in non-reciprocal systems as they naturally introduce two independent orders of symmetry. The first being the symmetry of the spins; through what angle do you have to rotate a spin for it to return to its original orientation. The second being any symmetry associated with the ``vision cone''; through what angle do you have to rotate in order to see what you originally saw. 

    We consider a lattice model of discrete spins each with orientation $\theta_i$. The order of symmetry of the spins is given by the positive integer $\mathfrak{p}$, hence spins with $\theta \pm n\pi/\mathfrak{p}$ are considered aligned for $n\in\mathbb{Z}$. Thus $\mathfrak{p} = 1,2,3$ denote polar, nematic and triatic spins, respectively. The dynamics of the spin at lattice site $i$ are governed by the stochastic equation
     \begin{equation}
        \label{eq:NRXY}
        \dot{\theta}_{i} \! =\!  J\sum_j K_{ij} \left[1\!  +\!  \sigma \cos \left( \mathfrak{q} \left( \theta_{i}\!  -\!  \alpha_{ij}  \right) \right) \right] \sin \left( \mathfrak{p} \left( \theta_{j}\!  -\!  \theta_{i} \right) \right) 
        +\eta_i \ .
    \end{equation}
     Here $J$ is an elastic constant and $\eta_i$ is uncorrelated Gaussian white noise with $\langle\eta_i(t)\eta_j(t')\rangle = 2D\delta(i-j)\delta(t-t')$. We have introduced the spatial kernel $K_{ij}$ that governs the range of local interactions. This is to avoid effects that arise from lattice anisotropy~\cite{dopierala2025inescapable,popli2025ordering}, see SI for details. 
    
    Non-reciprocity is introduced through a smooth angular kernel with magnitude $|\sigma|\leq1$~\cite{rouzaire2025nonreciprocal, rouzaire2026dynamics, popli2025ordering}. $\alpha_{ij}$ denotes the argument of the vector connecting lattice sites $i$ and $j$ and the positive integer $\mathfrak{q}$ denotes the order of symmetry of the non-reciprocity. For $\sigma=0$ we recover the standard XY model at symmetry $\mathfrak{p}$. When $\sigma\neq 0$, the response of spin $i$ to the state of spin $j$ is modulated according to the relative orientation of spin $i$ with respect to the vector connecting them. This is often expressed as the spins having ``vision cones'', in which they react more strongly to the lattice sites they are pointing toward. The introduction of $\mathfrak{q}$ allows this interaction to operate at higher orders of rotational symmetry. For example, $\mathfrak{q}=1$ denotes polar vision cones; the spins look in the direction they are pointing. $\mathfrak{q}=2$ indicates nematic vision cones, in which spins look both parallel and anti-parallel to the direction they are pointing; this would akin to equipping the spins with a rear-view mirror. In general, each spin takes more information from $\mathfrak{q}$ evenly spaced directions.

\begin{figure}[tb]
		\centering
		\includegraphics[width=\columnwidth]{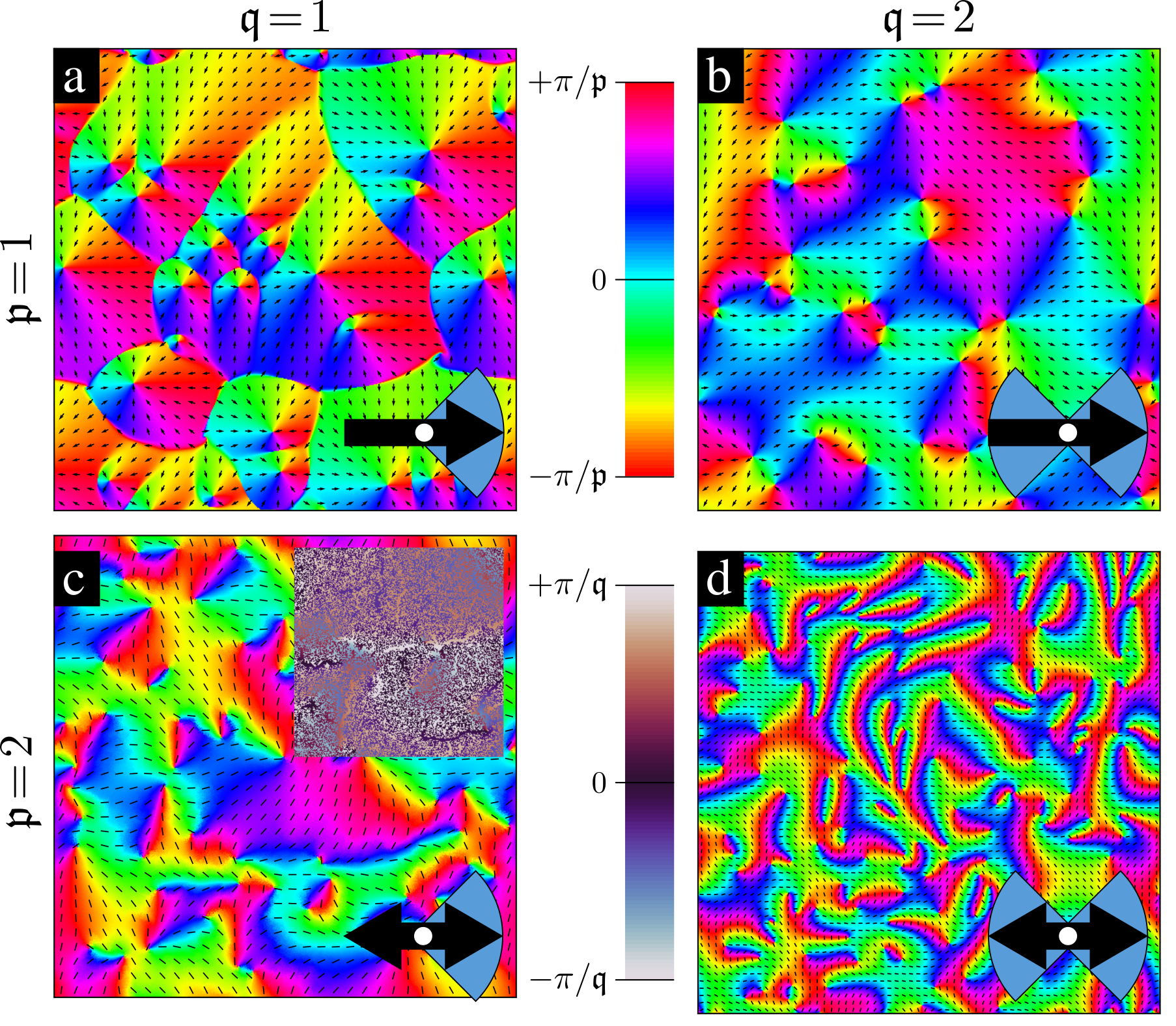}
		\caption{\label{fig:f1} Simulations of defect coarsening for a (a) $\{1,1\}$, (b) $\{1,2\}$, (c) $\{2,1\}$, (d) $\{2,2\}$ system. The bottom right of each panel shows the order of symmetry of the spin (black arrow) and the vision cones (blue sectors). (c)(inset) Shows the vision cone field for the $\{2,1\}$ system which displays local tracks of ordered regions. Color shows the local orientation of $\theta$, lines or arrows are added as a guide.}
    \end{figure}

Eq.~\ref{eq:NRXY} introduces two independent orders of symmetry, $\mathfrak{p}$ and $\mathfrak{q}$, governing elastic interactions and non-reciprocity, respectively; we denote each possible model by the pair $\{\mathfrak{p},\mathfrak{q}\}$. We numerically integrate Eq.~\ref{eq:NRXY} and visualize the spin field by plotting $\theta\mod{2\pi/\mathfrak{p}}$ with a cyclic color scale, see Fig.~\ref{fig:f1}. 
We first observe the systems evolving from a random configuration at low noise and high non-reciprocity; see SI for details. In all cases we see an initial formation of local order revealing topological defects of charge $\pm 1/\mathfrak{p}$ that gradually coarsen. 
    
The $\{1,1\}$ case is the standard Non-Reciprocal XY model, which has been studied at length~\cite{loos2023long, rouzaire2025nonreciprocal, rouzaire2026dynamics, popli2025ordering, bandini2025xy}. Non-reciprocity alters defect behavior, leading to a dramatic slowing down of coarsening~\cite{rouzaire2025nonreciprocal, vafa2022defect, besse2022metastability}. The slowly coarsening state is often referred to as a ``foam'' in which $+1$ defects form sinks, see Fig.~\ref{fig:f1}a. If we switch to nematic vision cones by simulating the $\{1,2\}$ system we no longer observe the dramatic slowdown, and the dynamics appear closer to that of the typical XY model. A key distinction is that almost all $+1$ defects are in either the source or sink configuration, but rarely vortex, see Fig.~\ref{fig:f1}b.

When we simulate the $\{2,2\}$ system we observe a significant change in defect behavior. The $+1/2$ topological defects begin to self-propel in a manner reminiscent of active nematics~\cite{giomi2014defect}, see Fig. ~\ref{fig:f1}c. We do not see defect proliferation in any of our simulations so a route to defect chaos is unlikely. At low noise, we often observe that after the defects have all annihilated, the system is left with a non-zero winding number ($>90\%$ of simulations). This topologically protected state is stable and stops the system reaching true global order. 
Similar states have been shown to be stable in the $\{1,1\}$ system though they do not arise naturally~\cite{rouzaire2026dynamics}, however it is possible in other active systems~\cite{rouzaire2025activity}.

Finally, we consider the $\{2,1\}$ case in which spins align nematically, but vision cones are polar. This is the first case in which $\mathfrak{q}/\mathfrak{p}\notin\mathbb{Z}$, which implies that for a given spin orientation there are multiple possible vision cone orientations. Thus two spins can be perfectly aligned but have anti-aligned vision cones. This is the same mixed symmetry associated with self propelled rods, polar biofilaments, and even people riding on kayaks~\cite{kayaks2024}.

Upon simulation of the $\{2,1\}$ case we see behavior similar to the $\{2,2\}$ system. The $+1/2$ defects self propel and at low noise defect coarsening often leaves the system with a non-zero winding number ($>90\%$ of simulations). At intermediate levels of noise, these states can spontaneously break symmetry and move, see SI Section II for details. 
        
In systems where $\mathfrak{q}/\mathfrak{p}\notin\mathbb{Z}$, we visualize the vision cone field by plotting $\theta\mod(2\pi/\mathfrak{q})$, see Fig.~\ref{fig:f1}c(inset). Here we see two important factors: First in the vast majority of the system there is no local alignment of vision cones, with adjacent vision cones often being $\pi$ out of phase. Second, we see channels of order in the vision code field which appear to originate from the cores of the $+1/2$ defects. It appears that as a defect self propels, it leaves a channel of ordered vision cones in it's wake.

    The most conspicuous effect on non-reciprocity is on the behavior of defects. In systems with $\mathfrak{p}=2$, $+1/2$ defects tend to self propel, a common feature of active nematic materials~\cite{giomi2014defect}. To assess defect self propulsion in more detail we simulate an isolated defect by initializing our simulation with a $\theta$ field given by the equation $\theta_i = (\phi_i + 2\pi l_i)/2$, where $\phi_i=\text{atan}(y_i/x_i)$ is the polar coordinate of lattice site $i$ around the center of the simulation where the defect is located. $l_i$ is a random integer that ensures that the vision cones are decorrelated when $\mathfrak{q}/\mathfrak{p}\notin\mathbb{Z}$.

\begin{figure}[htb]
		\centering
		\includegraphics[width=\columnwidth]{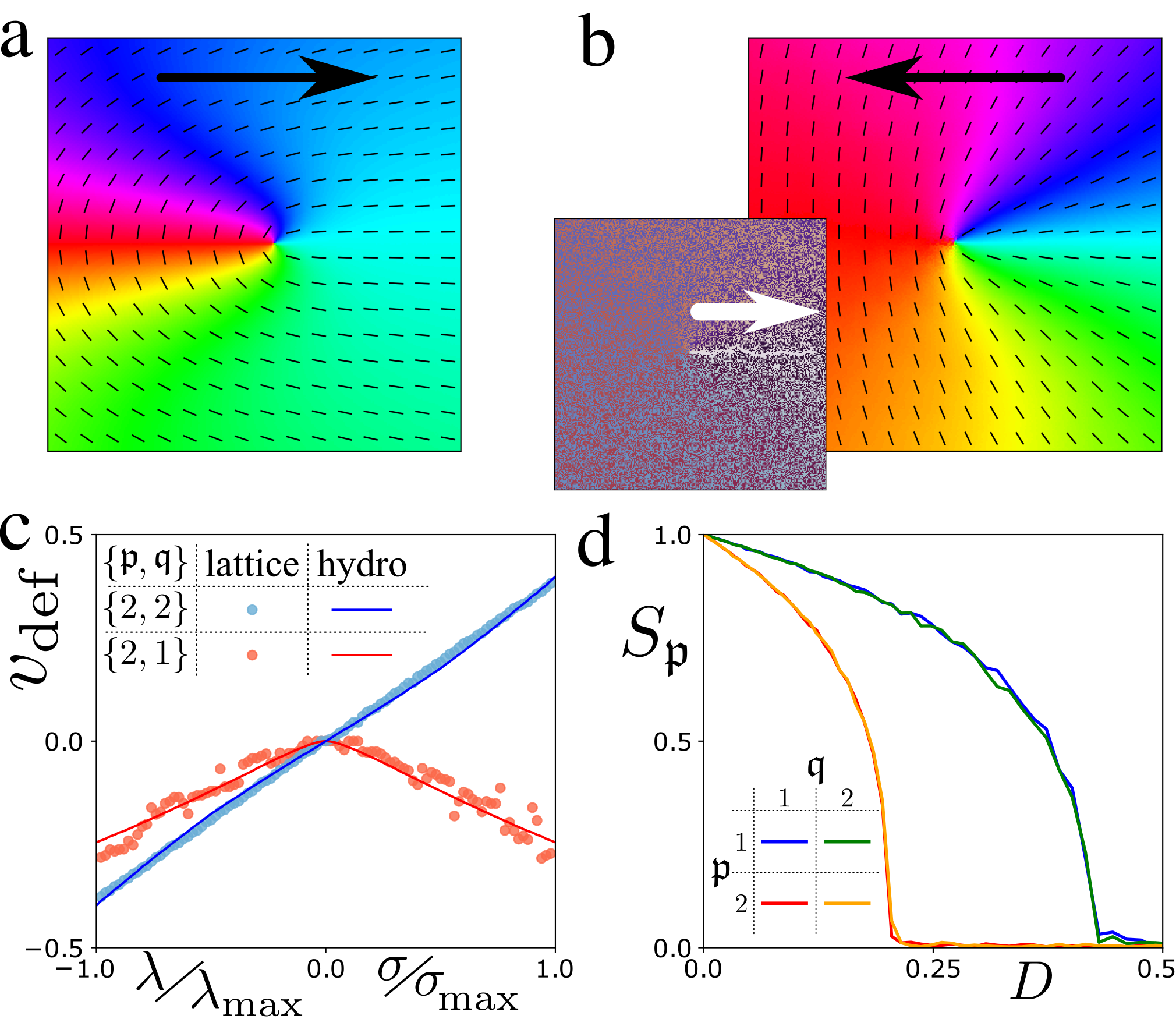}
		\caption{\label{fig:f2} (a)\&(b) Final shape of a self-propelling $+1/2$ defect in the (a) $\{2,2\}$ and (b) $\{2,1\}$ system for $\sigma>0$. Black arrows indicate the direction of motion. (b)(inset) The vision cone field for the $\{2,1\}$ system, the white arrow shows the orientation of the ordered track. (c) Defect self propulsion speed of an isolated $+1/2$ defect for the $\{2,2\}$ (red) and $\{2,1\}$ (blue) system at different levels of non-reciprocity. The points show the results of lattice based simulations, the lines show the results of hydrodynamic simulations. (d) Global order as a function of noise for the first four values of $\{\mathfrak{p},\mathfrak{q}\}$. }
    \end{figure}

    After an initial transient the $+1/2$ defects reach a stable shape, shown in Figs.~\ref{fig:f2}a\&b for the $\{2,2\}$ and $\{2,1\}$ systems respectively. In the $\{2,2\}$ system, the distortion is concentrated around the head of the defect where the spins are aligned azimuthally, see Fig.~\ref{fig:f2}a. Conversely, in the $\{2,1\}$ system distortion is concentrated at the tail where the spins are aligned radially, see Fig.~\ref{fig:f2}b. In both cases the defects self propel away from the distorted region (shown by the black arrows in Figs.~\ref{fig:f2}a,b) which is the opposite of passive systems~\cite{pearce_properties_2021}. Perhaps most strikingly, in the $\{2,1\}$ system we see that as the defect self propels, it leaves a trail of order in the vision cones pointing away from the core of the defect, Figs.~\ref{fig:f2}b (inset, white arrow).

    Changing the sign of $\sigma$ reverses the behavior of the $+1/2$ defect in the $\{2,2\}$ system, we observe self-propulsion toward the head, and distortions concentrated as splay around the tail. Conversely, in the $\{2,1\}$ system the only clear change for $\sigma<0$ is that the trail of order in the vision cone field now points toward the defect, see SI for details.
    
    Figure~\ref{fig:f2}c shows the velocity of an isolated $+1/2$ defect as a function of non-reciprocity $\sigma$ for both the $\{2,2\}$ and $\{2,1\}$ cases. In the $\{2,2\}$ system we observe a linear relationship between defect speed and $\sigma$ with the defect moving toward its tail (head) for $\sigma>0$ ($\sigma<0$). In the $\{2,1\}$ system the relationship appears linear in $|\sigma|$, with the defect moving toward its head independent of the sign of $\sigma$.
    
    The final feature we quantify is the effect of mixing symmetries on the order-disorder transition. We define the order parameter as
    $S_\mathfrak{p} = |\sum_{k}\exp(i\mathfrak{p}\,\theta_k)|/N$
    % \begin{equation}
    %     S_\mathfrak{p} = \frac{1}{N}\left|\sum_{k=1}^N\exp(i\mathfrak{p}\,\theta_k)\right|,
    % \end{equation}
    where the sum runs over the $N$ spins.
    %where $N$ is the number of spins in the system. 
    To avoid the previously mentioned topologically trapped states, we initialize the system with $\theta_i = 2\pi l_i/\mathfrak{p}$ with $l_i$ a random integer. This ensures initial order and zero topological defects, but also decorrelates the vision cone field.

    Fig.~\ref{fig:f2}d shows the global order parameter as a function of the noise, $D$, for the first four combinations of $\{\mathfrak{p},\mathfrak{q}\}$. We see a sharp transition to disorder as the noise amplitude is increased. The critical noise depends on the spin symmetry, $\mathfrak{p}$, but appears independent of the non-reciprocity symmetry, $\mathfrak{q}$. In the $\{2,1\}$ system we do not see any emergent vision cone alignment. If we initialize the system with aligned vision cones ($l_i=0$) we see them gradually decorrelate due to the effect of noise, see SI section II. Finally, we do not comment here on the nature of the phase transition or the existence of true long ranged order in the non-reciprocal XY model as it is not the focus of this manuscript. For insightful discussions on the topic we refer the reader to~\cite{loos2023long, dopierala2025inescapable, besse2022metastability, popli2025ordering, bandini2025xy}.

%    \section{Hydrodynamic model}

    To further understand how higher and mixed symmetries generate the phenomena we have reported, we coarse-grain the system and formulate a continuum model. We follow an approach similar to~\cite{fruchart2021non} which is fully outlined in SI section III. 

    As we have established, each simulation can contain different orders of orientational order. The $\{1,1\}$ and $\{1,2\}$ system are purely polar, the $\{2,2\}$ system is purely nematic, and the $\{2,1\}$ system can contain independent polar and nematic order simultaneously. In order to construct a hydrodynamic model, we introduce the order parameter for a field with $n$ fold rotational symmetry as 
    \begin{equation}
        f_n(\mathbf{r},t) = A_n(\mathbf{r},t)\exp(in\,\theta_n(\mathbf{r},t)).
    \end{equation}
    $A_n$ is the scalar order parameter which describes the local degree of order of the $n$th mode and $\theta_n$ describes the local orientation. Thus $n=1$ and $n=2$ are the polar and nematic order parameters, respectively. Since density is fixed in our model, $f_0=1$.

    During our coarse-graining we establish that $f_{\mathfrak{p}}$ is always the primary mode, which is able to spontaneously order provided we are below a critical noise. The critical noise is independent of $\mathfrak{q}$ and is given by $D_c\sim J/2\mathfrak{p}$, which is consistent with our observations, see Fig.~\ref{fig:f2}d. 
    In cases where $\mathfrak{q}/\mathfrak{p}\notin\mathbb{Z}$, we observe that it is possible for the $f_\q$ mode to become a secondary critical mode and true mixed symmetries arise. We will use our hydrodynamic description to understand the behavior of our simulations of Eq.~\ref{eq:NRXY}. 

%    \section{$\mathfrak{p} = \mathfrak{q}$}

    We focus first on the situation where $\mathfrak{p}=\mathfrak{q}$. In this case there is only one order parameter given by $f_\p$ which is governed by the equation
    \begin{multline}
        \label{eq:p=q}
        \partial_{t} f_{\mathfrak{p}} =  A(B-|f_{\mathfrak{p}}|^{2})]f_{\mathfrak{p}} + K \Delta f_{\mathfrak{p}}\\
        + \lambda_{1} f_\mathfrak{p}\left[ \partial^{\mathfrak{p}}_{-} f_{\mathfrak{p}}  -  \partial^{\mathfrak{p}}_{+} f_{-\mathfrak{p}}   \right] + \lambda_{2}f_{-\mathfrak{p}} \partial^{\mathfrak{p}}_{+} f_{\mathfrak{p}} %\\
    \end{multline}
    The terms in Eq.~\ref{eq:p=q} can be identified with three processes. The first terms govern the magnitude of $f_\mathfrak{p}$, describing the microscopic ordering of the system. $K$ plays the role of an elastic coefficient, penalizing gradients in the orientation field. Finally, the $\lambda$ terms are linear in $\sigma$ and represent the effect of non-reciprocity and hence are active in nature, i.e. cannot generally be derived from a free energy. 
    
\begin{figure}[htb]
		\centering
		\includegraphics[width=\columnwidth]{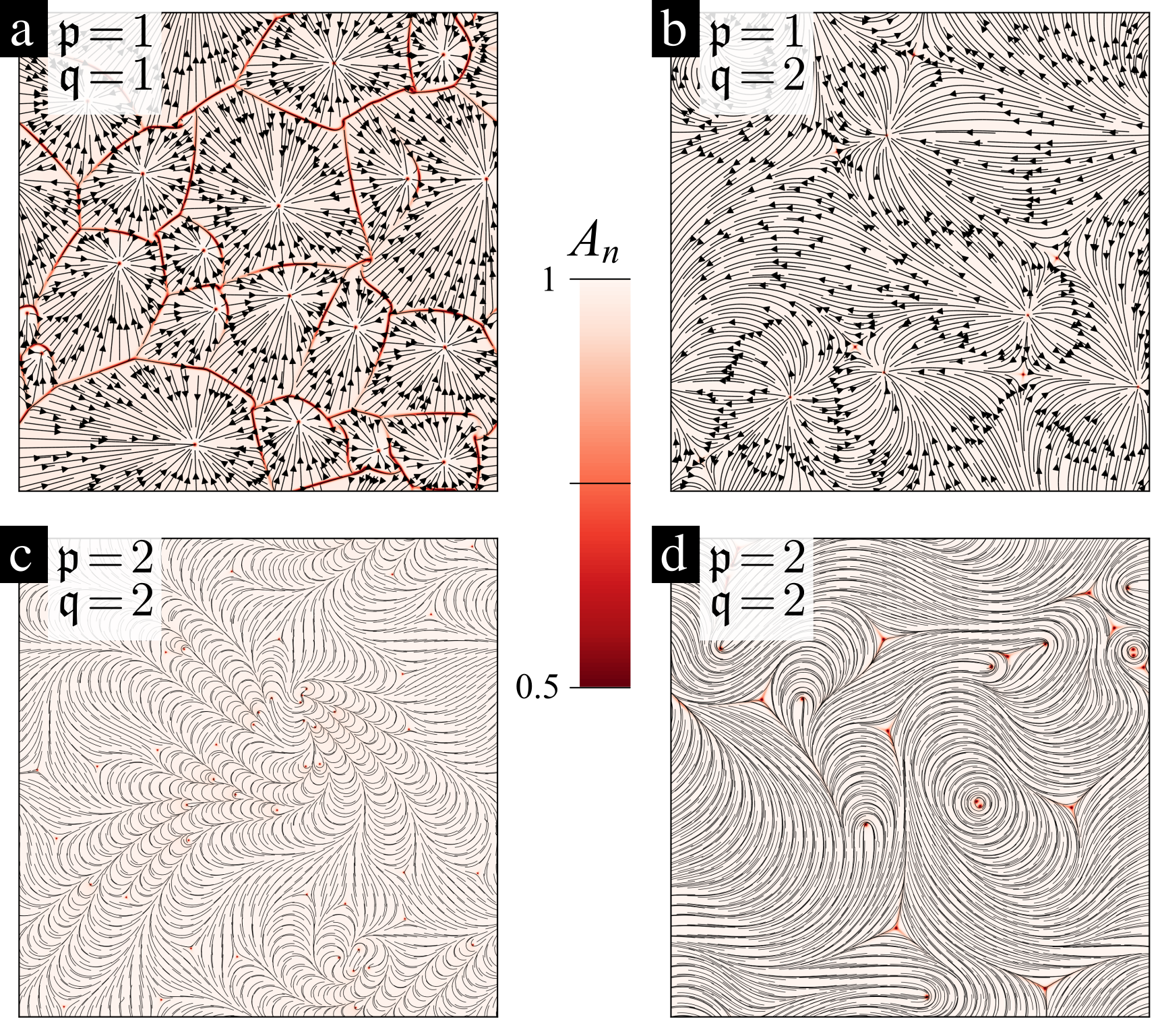}
		\caption{\label{fig:f3} Hydrodynamic simulations recreating the key features of the non-reciprocal XY models at different symmetries. (a) $\{1,1\}$, defect foam state, (b) $\{1,2\}$, rapidly coarsening system, and (c) $\{2,2\}$, self propelling $+1/2$ defect state. (d) Still in the $\{2,2\}$ system, it is possible to stabilize $+1$ defects by increasing the activity ($\lambda_2$). Color shows the magnitude of the order parameter and lines/arrows show the orientation.}
    \end{figure}

    If we set $\mathfrak{p}=1$, we can map the above equation directly onto the equation for Malthusian flocks~\cite{toner2012birth}; this is a previously known result for the $\{1,1\}$ case~\cite{dadhichi_nonmutual_2020}. A simulation of Eq.~\ref{eq:p=q} for $\mathfrak{p}=1$ gives very similar behavior to our lattice model, see Fig.~\ref{fig:f3}a. As in the lattice model, the defect foam state gradually coarsens into the ordered state. 
    
    Setting $\mathfrak{p}=2$ in Eq.~\ref{eq:p=q} gives us what we refer to as a non-reciprocal nematic which describes the behavior of the $\{2,2\}$ system at hydrodynamic scales. Simulating Eq.~\ref{eq:p=q} from random initial conditions recreates the major features of the $\{2,2\}$ system, being self propulsion of $+1/2$ defects that can eventually lead to a topologically trapped state, see Fig.~\ref{fig:f3}c. Simulating an isolated $+1/2$ defect reveals similar defect shape to the lattice model as well as a similar relationship between the defect speed and the non-reciprocity, see Fig.~\ref{fig:f2}c (red curve) and SI. 

    The hydrodynamic model sheds light on the mechanism of defect self propulsion. One action of the active terms of Eq.~\ref{eq:p=q} is to align the director field with variations in the scalar order parameter, see SI section V.  
    
    Around the core of a topological defect, gradients in the scalar order parameter are radial. When $\lambda_1>0$, the director aligns azimuthally, which causes the bend rich region of a $+1/2$ defect to expand. Eventually it is impossible to further expand the azimuthally aligned region of the nematic without the defect moving, which leads to self propulsion. Thus, the defect moves into aligned regions and leaves a wake of bend distortions, the exact opposite of the passive case. This causes the defect to often move away from nearby $-1/2$ defects, which can lead to it traversing the whole system before annihilating; this leaves the system with a non-zero winding number. By further increasing non-reciprocity, we can eventually stabilize $+1/2$ defects coming into close proximity, see Fig.~\ref{fig:f3}d. The defects approach tail to tail leading to vortex like structures; the $+1$ vortex is the only possible purely azimuthally aligned structure thus is very stable. 
    
    Finally, changing the sign of $\lambda_i$ in Eq.~\ref{eq:p=q} is equivalent to changing $f_\mathfrak{p}\rightarrow-f_\mathfrak{p}$, which is a rotation of the director field by $\pi/\mathfrak{p}$. Around a $+1/2$ defect, this transformation exchanges splay for bend and reverses the orientation of the defect, leading to motion toward a splay dominated tail; exactly what we see for negative $\sigma$ in Fig.~\ref{fig:f2}a,c and SI.

    The $\{1,2\}$ system is in the class in which $\mathfrak{q}/\mathfrak{p}\in\mathbb{Z}$, thus can be described by a single order parameter, the dynamics of which are given by
    \begin{equation}\label{eq:q=2p}
        \begin{aligned}
            \partial_{t} f_1 = &\ A(B-|f_1|^2)f_1 + K \Delta  f_1 -  \lambda_1 \partial_+^{2}f_{-1}, 
        \end{aligned}
    \end{equation}
    which is simulated in Fig.~\ref{fig:f3}b. Interestingly, this dynamical equation can be mapped to the relaxation of a polar field with different elastic coefficients for bend and splay deformations controlled by $\lambda_1$, see SI section VI. 
    Thus at a coarse-grained level we can say that it behaves like an equilibrium system, a scenario already reported in multi-species non-reciprocal systems \cite{dinelli2023non}.

\begin{figure}[tb]
		\centering
		\includegraphics[width=\columnwidth]{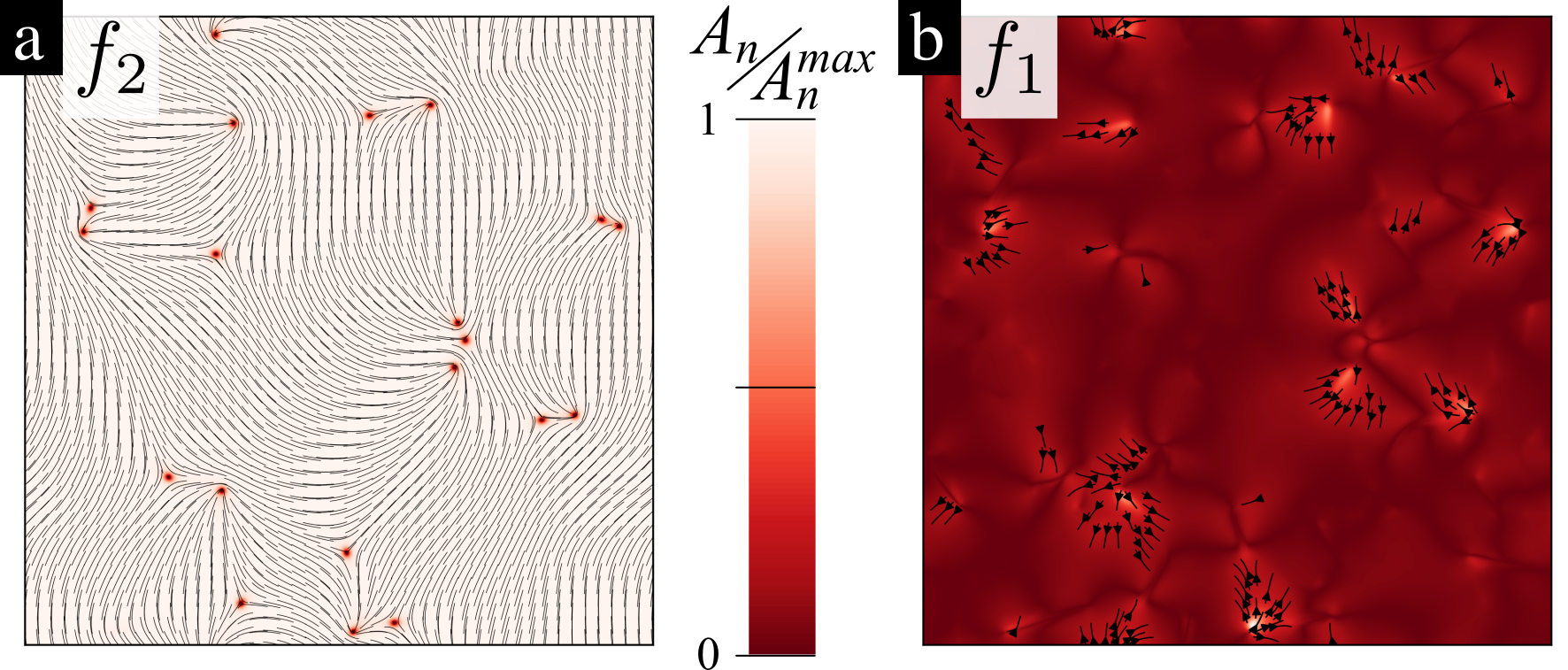}
		\caption{\label{fig:f4} Hydrodynamic simulation of the $\{2,1\}$ system. (a) The $f_2$ primary mode and (b) the $f_1$ secondary mode.} 
    \end{figure}

    In the final section of the manuscript, we turn our attention to the $\{2,1\}$ system, in which $\mathfrak{q}/\mathfrak{p}\notin\mathbb{Z}$; thus we expect the spin and vision cone fields to be independent and there is true mixed symmetry. The nematic spin field is the primary critical mode described by $f_2$, the polar vision cone field is a secondary critical mode described by $f_1$. Their dynamics follow
    \begin{subequations}
        \label{eq:21}
        \begin{align}
        \partial_{t} f_2 = &\ A(B-|f_2|^2)f_2 + K   \Delta  f_2\notag \\
        &+ \lambda_1\left[ f_1 \cdot \partial_{-} f_2  + f_{-1} \cdot \partial_{+} f_{2}\right]- \lambda_2\partial_{+} f_{2}\cdot \partial_{+} f_{-2},\label{eq:21_2}\\
        \partial_{t} f_{1} = & \ C_1f_{2}\cdot f_{-1}- C_2f_{1} - C_3|f_{2}|^2f_{1}  + \frac{K}{2} f_{-1}\Delta f_{2}\notag\\
        & + \frac{\lambda_1}{2}\left[ \partial_{-} f_{2} + f_{-2} \cdot \partial_{+} f_{2} - f_{2} \cdot \partial_{+} f_{-2} \right].\label{eq:21_1}
        \end{align}    
    \end{subequations}
    The $C_i$ parameters control relaxation of the $f_1$ field, which is now coupled to the $f_2$ field. The fields are also coupled through active terms controlled by the $\lambda_i$ parameters which arise from the action of non-reciprocity. Simulating Eqs.~\ref{eq:21} leads to a recreation of the $\{2,1\}$ system which captures the key features observed in the lattice model: self propulsion of $+1/2$ defects which generate local order in $f_1$, see Fig.~\ref{fig:f4}a,b. Simulating an isolated $+1/2$ defect again recreates the defect shape and motion of the lattice model, see Fig.~\ref{fig:f2}c (blue line) and SI.
    
    Once again, our hydrodynamic model sheds light on the self propulsion mechanism of $+1/2$ defects in the $\{2,1\}$ system. The leading order active contribution to Eq.~\ref{eq:21_1} has the form $\partial^{\mathfrak{q}}_{-} f_{\mathfrak{p}}$ governed by $\lambda_1$, this causes $f_1$ to align with gradients of $f_2$. In the $\{2,1\}$ system, this results in a strong aligning force at the center of a $+1/2$ defect causing $f_1$ to point toward its tail (head) for $\lambda_1>0$ ($\lambda_1<0$), as we observe in the microscopic model. 
    
    The leading order active term in Eq.~\ref{eq:21_2} is also governed by $\lambda_1$. This has the form of an advective term in which $f_2$ is advected antiparallel (parallel) to the $f_1$ field for $\lambda_1>0$ ($\lambda_1<0$). The polar vision cones locally bias the flow of information through the system, leading to a net information flux~\cite{rouzaire2026dynamics}. Since $f_1$ is ordered at the core of a $+1/2$ defect, this results in local advection of the defect and thus self propulsion. Since both alignment of $f_1$ and advection of $f_2$ depend on $\lambda_1$, we expect a quadratic relationship between defect speed and non-reciprocity; this is true for $\lambda_1$ small. When $\lambda_1$ is increased, the magnitude of $f_1$ saturates and the defect speed rather scales with $|\lambda_1|$, mirroring what we observe in Fig.~\ref{fig:f2}c. Reversing $\lambda_1$ reverses both the alignment of $f_1$ and the direction of advection, leaving defect self propulsion unaffected. This mechanism of defect self propulsion mirrors that observed in active nematics; the $\lambda_1$ term in Eq.~\ref{eq:21_1} has identical form to the active force generation in an active nematic. This generates a flow parallel to the defect which then advects the director field in an identical fashion to the $\lambda_1$ term in Eq.~\ref{eq:21_2}.

    In this manuscript we have shown that when different orders of rotational symmetry are combined in active non-reciprocal systems, a wide variety of phenomena can occur. This includes altering the stability of different states, defect self propulsion, and multiple simultaneous order parameters. 

    We demonstrate that changing the orders of symmetry of the microscopic processes driving a system can have a great effect on the eventual coarse-grained behavior. Indeed, of the four cases studied here, only three remain active at the coarse-grained level. 

    While our results obviously increase our understanding of non-reciprocal and active systems, we expect them to have a large impact on the description of biological tissues. Biological tissues are by their nature out of equilibrium, and are regularly described using active and non-reciprocal models. They regularly feature complex environments in which many symmetries are combined simultaneously. Finally, and most importantly, biological tissues are capable of information processing and decision making. Cells sense their environment and react accordingly, a process that opens the door to non-reciprocal interactions. As such, we expect our findings to be essential in building a full understanding of cellular organization at the tissue scale.

	\begin{acknowledgments}
        The authors acknowledge Demian Levis and Ignacio Pagonabarraga for insightful discussions.
		YA \& DJGP acknowledge funding from the Swiss National Science Foundation under starting grant TMSGI2 211367. YR acknowledges Ministerio de Ciencia, Innovaci\'on y Universidades MCIU/AEI/FEDER for financial support under grant agreement PID2024-156516NB-100 AEI/FEDER-EU.
	\end{acknowledgments}

    \bibliography{bibfile}% Produces the bibliography via BibTeX.
    \FloatBarrier

    \onecolumngrid

\section{Numerical Methods}

\subsection{Lattice simulations}

Data is generated by GPU accelerated simulations written in CUDA. 

For the lattice model, Eq.1 (Main text) is simulated on a square lattice. Numerical integration is performed using Euler's method. The parameters used are given in the table \ref{tab:lattice}. 

To avoid lattice effects, we employ an isotropic Gaussian kernel given by: 
\begin{equation}
	K_{ij} = \begin{cases}
		A\exp\left(-\frac{\left(\underline{R}_i - \underline{R}_j\right)^2}{2\kappa^2}\right), & \text{if } i\neq j \\
		0, & \text{if } i=j
	\end{cases}
\end{equation}
Here $\underline{R}_i$ is the position of lattice site $i$ and $A$ is a normalization constant given by:
\begin{equation}
	A^{-1} = \sum_{i,j}K_{ij}.
\end{equation}
We choose $\kappa = 3/\sqrt{2}$ in units of lattice spacing. The kernel is truncated when it reaches $5\%$ of it maximum magnitude. This results in a kernel of $\approx100$ lattice sites which significantly negates lattice anisotropy effects while maintaining computational efficiency.

\begin{figure}[h!]
	\centering
	\includegraphics[width=0.75\columnwidth]{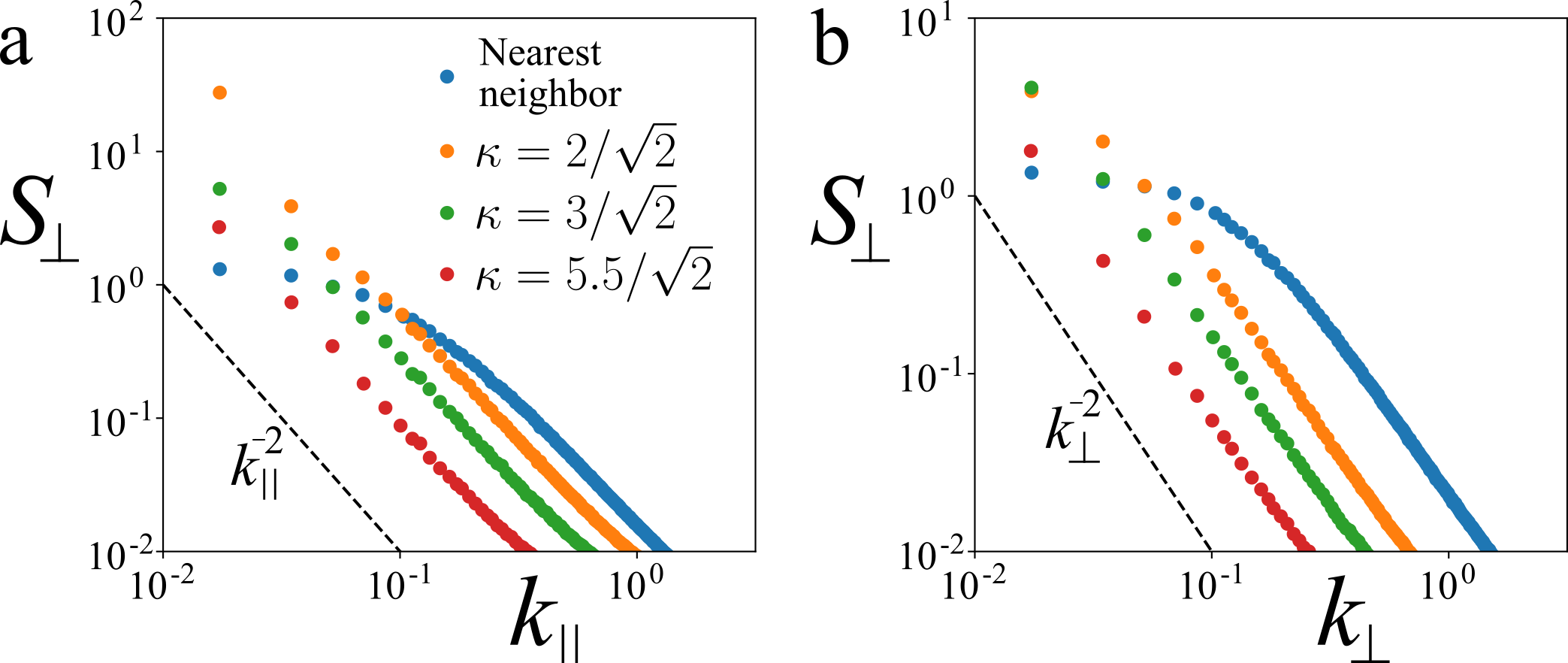}
	\caption{\label{fig:SF}  } 
\end{figure}

It was recently  highlighted that nonreciprocal XY models defined on discrete lattices can retain lattice-induced anisotropy at hydrodynamic scales~\cite{ dopierala2025inescapable, popli2025ordering}. This can generate an effective mass for orientational fluctuations, manifested as a saturation of the transverse structure factor at small wavenumbers. To suppress these effects, we employ a radially symmetric Gaussian interaction kernel that extends the coupling over many lattice sites, thereby reducing the influence of preferred lattice directions. In Fig.\ref{fig:SF}, we compare the structure factors $S_\perp(k_\parallel)$ and $S_\perp(k_\perp)$ for nearest-neighbor interactions and Gaussian kernels of increasing width. Whereas nearest-neighbor interactions exhibit a flattening at small wavenumbers, this feature is strongly suppressed for broader kernels, with no comparable saturation resolved over the accessible range. This suggests that lattice-induced anisotropy is sufficiently weak to be neglected for the system sizes and length scales considered here, allowing us to compare the microscopic dynamics with a rotationally invariant hydrodynamic description. This does not imply an exact restoration of continuous rotational symmetry at arbitrarily large scales for the lattice simulations.% We expect the effects could be further mitigated by employing a randomized lattice, though we do not use that here. 

\begin{table}[]
	\centering
	\begin{tabular}{c|c|c|c|c|c|c}
		Figure & Grid size & Time step $\Delta t$  & Boundaries & Elasticity $J$ & Noise $D$ & Non-reciprocity $\sigma$ \\
		Fig. 1 (Main text) & $512$ & $10^{-2}$ & Periodic & $1$ & $0.05$ & $0.8$\\
		Fig. 2d (Main text) & $512$ & $10^{-2}$ & Periodic & $1$ & See Figure & $0.64$\\
		Fig. 2c (Main text) & $320$ & $4\times 10^{-5}$ & Neumann & $1$ & $0.0160$ & $0.64$ (max)\\
		Fig. 2a,b (Main text) & $2048$ & $10^{-2}$ & Neumann & $1$ & $0$ & $0.8$\\
		Fig.~\ref{fig:SF} (SI) & $512$ & $10^{-2}$ & Periodic & $1$ & $0.20$ & $2.5$\\
		Fig.~\ref{fig:neg_sigma_p2} (SI) & $2048$ & $10^{-2}$ & Neumann & $1$ & $0$ & $0.8$\\
		Fig.~\ref{fig:decorrelation} (SI) & $128$ & $10^{-2}$ & Periodic & $1$ & See Figure & $0.80$\\
		Fig.~\ref{fig:neg_sigma_p1} (SI) & $512$ & $10^{-2}$ & Periodic & $1$ & $0.05$ & $0.8$\\
		Fig.~\ref{fig:kymo} (SI) & $320$ & $10^{-2}$ & Periodic & $1$ & See Figure & $0.64$\\
	\end{tabular}
	\caption{Simulation parameters: Lattice model}
	\label{tab:lattice}
\end{table}

\FloatBarrier
\subsection{Hydrodynamic simulations}

All hydrodynamic simulations are generated using finite difference methods on a square lattice. Numerical integration is performed using Euler's method. The parameters used are given in tables \ref{tab:hydro_1},\ref{tab:hydro_2}.

The real and imaginary parts of each field can be translated into a more recognizable format using the following identities. 

\begin{equation}
	\begin{aligned}
		f_1 &= A_1\exp[i\theta] =  P_x + iP_y\\
		f_2 &= A_2\exp[2i\theta] =  Q_{xx} + iQ_{xy}
	\end{aligned}
\end{equation}

In the polar case $\underline{P} = [P_x,P_y]$ is a vector aligned with the local orientation of the spins, $\theta$ and magnitude $A_1$.  

In the nematic case the symmetric, traceless tensor $\underline{\underline{Q}}$ describes the local nematic orientation and scalar order. The largest eigenvalue of $\underline{\underline{Q}}$ is $A_2$ and the associated eigenvector is parallel to the local nematic director. 

The equations used to simulate the hydrodynamic fields are included below in their standard vector/tensor form. 
\begin{itemize}
	
	\item{Standard Notation Equations $\mathfrak{p}=1$, $\mathfrak{q}=1$}
	\begin{equation}
		\begin{aligned}
			\partial_{t} P_{i} = & \ A\left(B-P_{j}P_{j}\right) P_{i} + K \nabla^{2}P_{i} \\
			& + \left(2 \lambda_{1} + \lambda_{2} \right) P_{j} \partial_{j} P_{i} \\
			& + \left( \lambda_{2} - 2\lambda_{1}\right) \partial_{i} \left(\frac{P_{j}P_{j}}{2} \right)\\
			& - \lambda_{2} \left(\partial_{j} P_{j}\right) P_{i}
		\end{aligned}
	\end{equation}
	\item{Standard Notation Equations $\mathfrak{p}=1$, $\mathfrak{q}=2$}
	\begin{equation}
		\begin{aligned}
			\partial_{t} P_{i} = &\  A\left(B-P_{j}P_{j}\right) P_{i} + (K+\lambda_{1}) \nabla^{2}P_{i} - 2\lambda_{1} \partial_{i} \partial_{j} P_{j}
		\end{aligned}
	\end{equation}
	\item{Standard Notation Equations $\mathfrak{p}=2$, $\mathfrak{q}=2$}
	\begin{equation}
		\begin{aligned}
			\partial_{t} Q_{ij} = & \ A\left(B-2Q_{kl}Q_{kl}\right) Q_{ij} + K \nabla^{2}Q_{ij} \\
			& + \lambda_{1} \left[ Q_{ik} \partial_{k} \partial_{l} Q_{jl} + Q_{jk}\partial_{k}\partial_{l} Q_{il} - Q_{ik} \partial_{j} \partial_{l} Q_{kl} - Q_{jk} \partial_{i} \partial_{l} Q_{kl} \right]  \\
			& + \lambda_{2} \left[ 2Q_{kl} \partial_{k} \partial_{l} Q_{ij} - Q_{ik}\partial_{k} \partial_{l} Q_{jl} - Q_{jk} \partial_{k} \partial_{l} Q_{il} + \delta_{ij} Q_{kl}\partial_{k}\partial_{m} Q_{lm} \right]
		\end{aligned}
	\end{equation}
	\item{Standard Notation Equations $\mathfrak{p}=2$, $\mathfrak{q}=1$}
	\begin{equation}
		\begin{aligned}
			\partial_{t} Q_{ij} = & \ A\left(B-2Q_{kl}Q_{kl}\right) Q_{ij} + K \nabla^{2}Q_{ij} \\
			& + 2 \lambda_{1} P_{k} \partial_{k} Q_{ij} \\
			& - \lambda_{2} \left[ (\partial_{i} Q_{kl}) (\partial_{j} Q_{kl}) - \frac{\delta_{ij}}{2} (\partial_{m} Q_{kl} ) (\partial_{m} Q_{kl})\right]
		\end{aligned}
	\end{equation}
	\begin{equation}
		\begin{aligned}
			\partial_{t} P_{i} = & \ C_{1} Q_{ij} P_{j} - C_{2} P_{i} - C_{3} \left( Q_{kl} Q_{kl}\right) P_{i} + \frac{K}{2} P_{j} \nabla^{2} Q_{ij} \\
			& + \frac{\lambda_{1}}{2} \left[ \partial_{j}Q_{ij} + Q_{jk} \partial_{j} Q_{ik} - Q_{ik}\partial_{j} Q_{jk}\right]
		\end{aligned}
	\end{equation}
\end{itemize}

\begin{table}[]
	\centering
	\begin{tabular}{c|c|c|c|c|c|c|c|c|c|c}
		Figure & $N$ & $\Delta t$ & BC & $\Delta x$ & $K$ & $A$ & $B$ & $\lambda_1$ & $\lambda_2$\\
		Fig. 2c (Main text) $\{2,2\}$ & $512$ & $2.5\times 10^{-3}$ & Neumann & $1$ & $2.5$ & $1$ & $1$ & $0.84$ (max) & $0.756$ (max)\\
		Fig. 3a (Main text) & $1024$ & $5\times10^{-3}$ & Periodic & $0.5$ & $1$ & $1$ & $1$ & $0.64$ & $0.48$\\
		Fig. 3b (Main text) & $1024$ & $5\times10^{-3}$ & Periodic & $0.5$ & $1$ & $1$ & $1$ & $0.64$ & NA\\
		Fig. 3c (Main text) & $1024$ & $5\times10^{-3}$ & Periodic & $0.5$ & $1$ & $1$ & $1$ & $0.34$ & $0.24$\\
		Fig. 3d (Main text) & $1024$ & $5\times10^{-3}$ & Periodic & $0.5$ & $1$ & $1$ & $1$ & $0.34$ & $0.30$\\
		Fig.~\ref{fig:S5a} & $2048$ & $8\times10^{-4}$ & Neumann & $0.125$ & $2.5$ & $1$ & $1$ & $\pm0.84$ & $\pm 0.76$\\
	\end{tabular}
	\caption{Simulation parameters: Hydrodynamic model ($\mathfrak{q}/\mathfrak{p}\in\mathbb{Z}$)}
	\label{tab:hydro_1}
\end{table}

\begin{table}[]
	\centering
	\begin{tabular}{c|c|c|c|c|c|c|c|c|c|c|c|c|c}
		Figure & $N$ & $\Delta t$ & BC & $\Delta x$ & $K$ & $A$ & $B$ & $\lambda_1$ & $\lambda_2$ & $C_1$ & $C_2$ & $C_3$\\
		Fig. 2c (Main text) $\{2,1\}$ & $512$ & $2.5\times 10^{-3}$ & Neumann & $2.5$ & $1$ & $1$ & $1$ & $0.2$ (max)&  $0$ (max) & $0.054$ & $0.0275$ & $0.0275$\\
		Fig. 4 (Main text) $\{2,1\}$ & $1024$ & $5\times 10^{-3} $ & Periodic & $0.5$ & $1$ & $1$ & $1$ & $0.16$ & $0$ & $0.054$ & $0.0275$ & $0.0275$\\
		Fig.~\ref{fig:S5b} & $2048$ & $8\times 10^{-4}$ & Neumann & $0.125$ & $2.5$ & $1$ & $1$ & $0.28$ & $0.0$ & $0.054$ & $0.0275$ & $0.0275$\\
	\end{tabular}
	\caption{Simulation parameters: Hydrodynamic model ($\mathfrak{q}/\mathfrak{p}\notin\mathbb{Z}$)}
	\label{tab:hydro_2}
\end{table}

\subsection*{Data availability}
Source data are provided with this paper. The numerical data required to recreate all plots from the main text and supplementary information is provided in a public GitHub repository \cite{github}.
The source data file is stored as tab delineated text files (.txt). Upon extraction, the source data file contains a folder for each figure in the main text and supplementary information. Within each folder is all the data required to completely remake the figure. For convenience, a python script is included which, when run, is able to reconstruct all the plots as they appear in the manuscript.

Source code to recreate a representative simulation is also provided at the same GitHub repository \cite{github}.

\section{Supplementary Numerical Results}

\subsection{Lattice model}

\subsubsection{Defect shape and motion for $\sigma<0$}

\begin{figure}[h]
	\centering
	\includegraphics[width=\columnwidth]{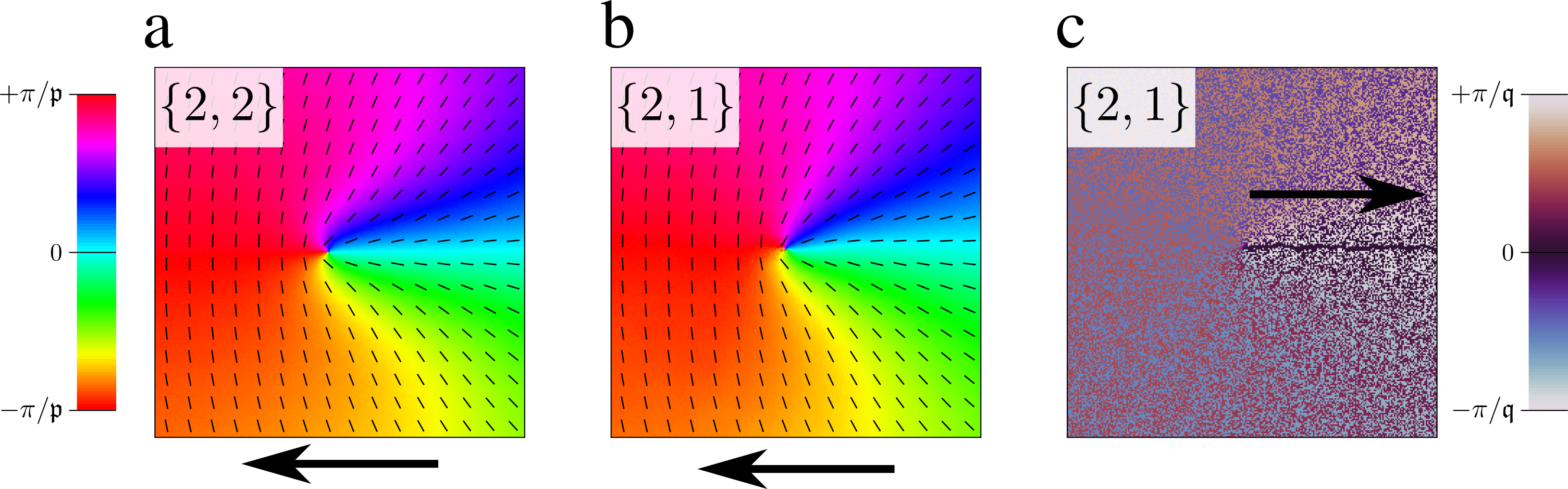}
	\caption{\label{fig:neg_sigma_p2} (a) Shape of an isolated $+1/2$ defect in a $\{2,2\}$ system with $\sigma<0$. The defect has a splay dominated configuration and moves toward its head. This is the opposite of when $\sigma>0$. (b) Shape of an isolated $+1/2$ defect in a $\{2,1\}$ system with $\sigma<0$. The defect has a splay dominated configuration and moves toward its head, which is identical to when $\sigma>0$. (c) The defect generates local alignment in the vision cones which point to the tail of the defect, this is the opposite of when $\sigma>0$. } 
\end{figure}

\FloatBarrier
%\subsubsection{Defect speed with correlated vision cones in 2,1 model}

%\dan{Recreate Fig.2b with correlated vision cones in the 2,1 model. Does not need so many values of sigma. Perhaps show with the grain boundary in the two symmetric positions (head or tail).}

\subsubsection{Decorrelation time of vision cones in $\{2,1\}$ model}

We consider a lattice simulation of the $\{2,1\}$ system in a super-ordered state ($\theta_i\approx\theta_0$), with low noise $D$. This has all vision cones perfectly aligned. 

To first approximation, the deterministic parts of the lattice equation generally act to suppress fluctuations in this situation. Thus the only way for a vision cone to flip, is to have $|\eta_i|\in(\pi/2,3\pi/2)$, which happens with probability 
\begin{equation}
	P_{\mathrm{flip}}\approx 2\sqrt{4\pi D}\int_{\pi/2}^\infty \exp(-x^2/4D)\mathrm{d}x.
\end{equation}

Thus we expect an independent flipping rate of the vision cones proportional to $\omega_{\mathrm{flip}}\sim P_{\mathrm{flip}}^{-1}$. The probability of a vision cone to perform an even number of flips is given by 
\begin{equation}
	p(t) = (1+\exp(-2\omega_{\mathrm{flip}}t))/2.
\end{equation}
Thus the expected polar order parameter can be written simply as
\begin{equation}
	S_1 = \exp(-2\omega_{\mathrm{flip}}t).
\end{equation}
In the long time limit we expect fluctuations around $S_1=0$ of the order of $N^{-0.5}$.

\begin{figure}[h]
	\centering
	\includegraphics[width=0.8\columnwidth]{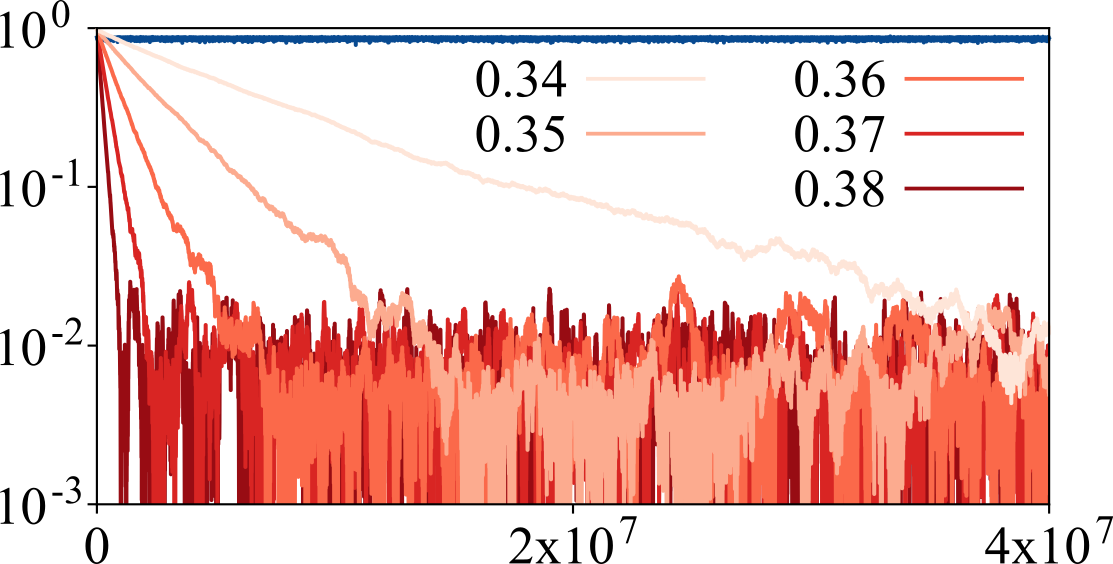}
	\caption{\label{fig:decorrelation} Evolution of polar order parameter $S_1$ for different noise values starting from an initially perfectly ordered state. We observe a clear exponential decay until fluctuating around $S_1\approx10^{-2}$. The nematic order parameter $S_2$ is stable (shown for $D=0.38$ $S_2\approx0.86$) indicating nematic order.} 
\end{figure}

We plot the evolution of $S_1$ starting from a super ordered state in Fig.~\ref{fig:decorrelation} for a system of $N=128^2$ spins for a range of noise values below the critical noise. We see a clear exponential decay with a rate that increases with $D$ as expected. In the long time limit, the fluctuations are of the order $\approx 128^{-1}$ as expected. Thus we don't see any suggestion of the dynamics creating order in the vision cones in an aligned state. 

\FloatBarrier
\subsubsection{Negative $\sigma$ for $\{1,1\}$ and $\{1,2\}$ systems}

\begin{figure}[h]
	\centering
	\includegraphics[width=\columnwidth]{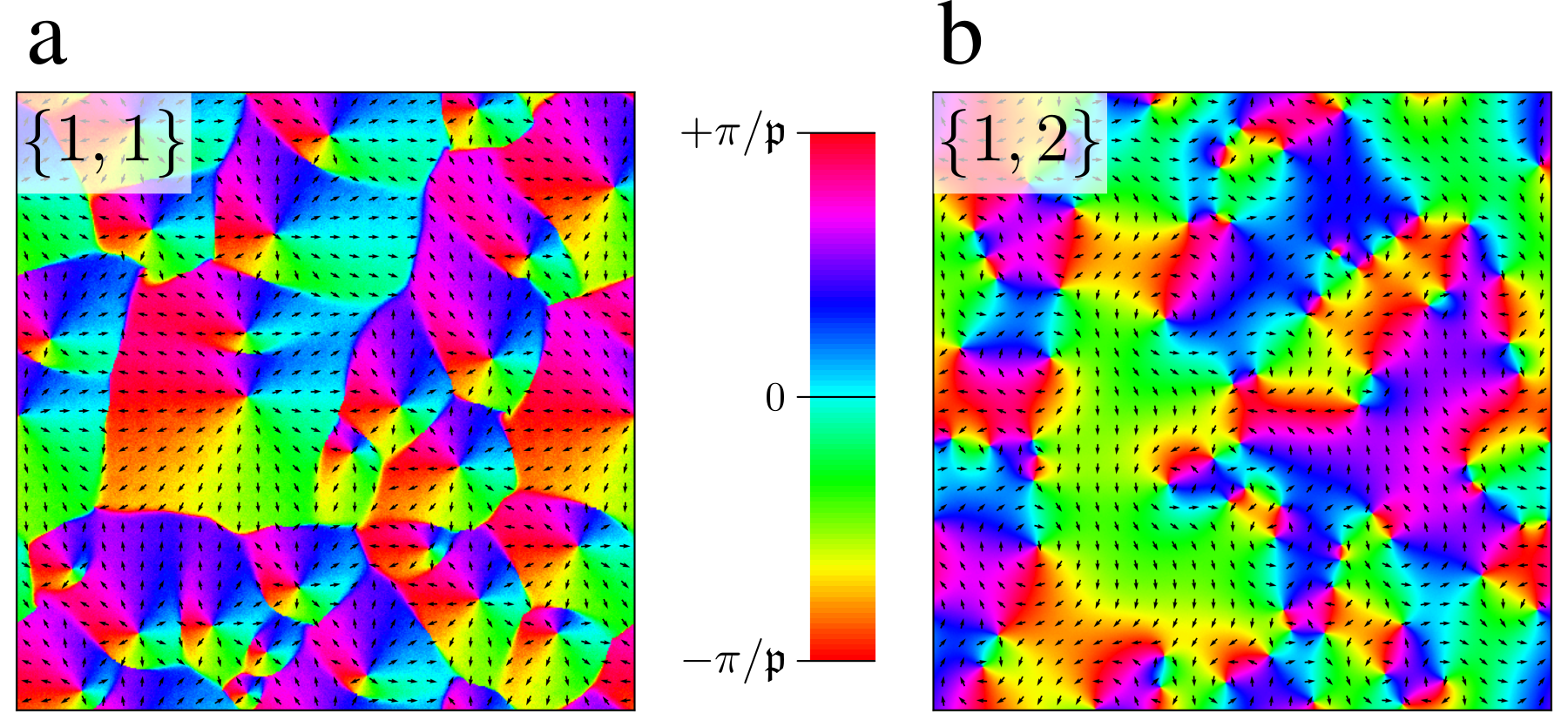}
	\caption{\label{fig:neg_sigma_p1} (a) Typical snapshot of defect coarsening in the (a) $\{1,1\}$ and (b) $\{1,2\}$ system with $\sigma<0$.} 
\end{figure}

The behavior for negative $\sigma$ can be understood by considering the hydrodynamic model. The $\{1,1\}$ model is governed by Eq.~5~(main text), and changing the sign of $\sigma$ is equivalent to changing the sign of $\lambda_i$. This is equivalent to changing $f_1\rightarrow -f_1$, which represents a rotation by $\pi$. Hence in Fig.~\ref{fig:neg_sigma_p1}a the stable $+1$ defect configuration in the foam state becomes a source instead of a sink. 

The hydrodynamic model for the $\{1,2\}$ system (Eq.~6~(main text)) has potential dynamics (shown below) in which $\lambda_1$ distinguishes between splay and bend distortions. A sign change of $\sigma$ is represented by a sign change of $\lambda_1$, which switches between bend and splay being the favored distortion. Hence in Fig.~\ref{fig:neg_sigma_p1}b the stable $+1$ defect configuration during coarsening is a vortex which maximizes bend. 

\FloatBarrier
\subsubsection{Traveling wave states for $\{2,1\}$ system.}

\begin{figure}[h]
	\centering
	\includegraphics[width=0.99\columnwidth]{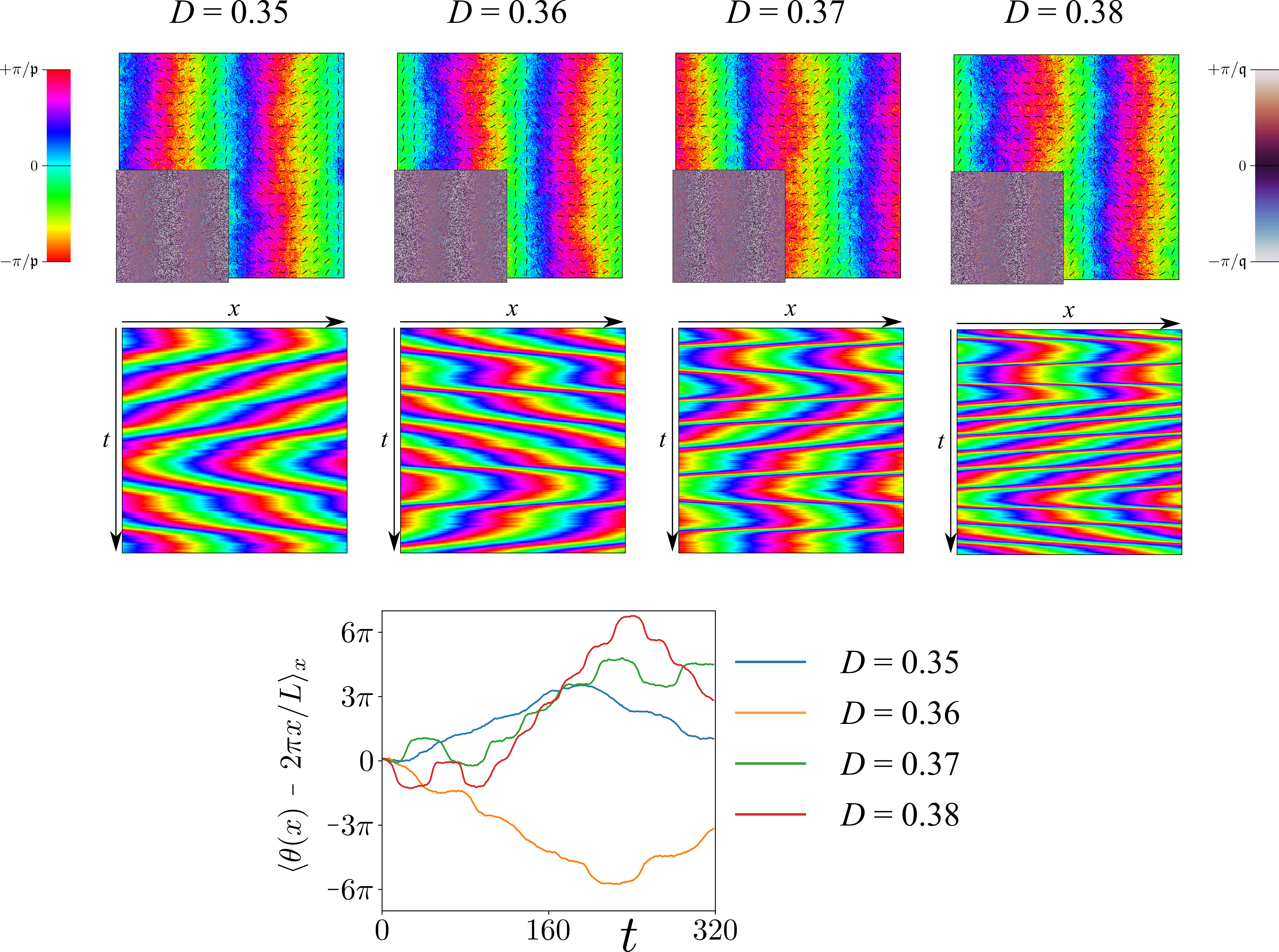}
	\caption{\label{fig:kymo} Topologically trapped states that self propel. (Top row) Snapshot of the spin and vision cone fields at the end of a simulation of a system with a winding number. The noise is increased from left to right. (Middle row) Kymographs showing the motion of the topologically trapped states over time. (Bottom) Graph showing the evolution of the global phase of the pattern over time.} 
\end{figure}

When simulated from random initial conditions, the most likely outcome for the $\{2,1\}$ system below critical noise is to arrive at a state with a global winding number; we refer to this as the topologically trapped state. Close to the order-disorder transition, we observe spontaneous translation of these topologically trapped states. This is visualized in the kymographs in Fig.~\ref{fig:kymo}. 

We can quantify the traveling states by defining a global phase for the topologically trapped state. This is given by
\begin{equation}
	\langle\theta (\underline{x}) - \underline{q}.\underline{x}\rangle.
\end{equation}
Here $\theta(\underline{x})$ is the spin orientation at position $\underline{x}$ and $\underline{q}$ is the wave vector associated with the winding number. In Fig.~\ref{fig:kymo} the topological winding is in the $\hat{x}$ direction and has winding number $2$; thus $\underline{q} = 2\pi\underline{\hat{x}}/L$. 

The global phase is plotted as a function of time for the four values of noise at the bottom of Fig.~\ref{fig:kymo}. The phase appears to move in jumps of $\pm\pi$, which become faster and more distinct as noise is increased. 

\FloatBarrier
\subsection{Hydrodynamic model}

\subsubsection{Defect shape and speed}

\begin{figure}[h]
	\centering
	\includegraphics[width=0.5\columnwidth]{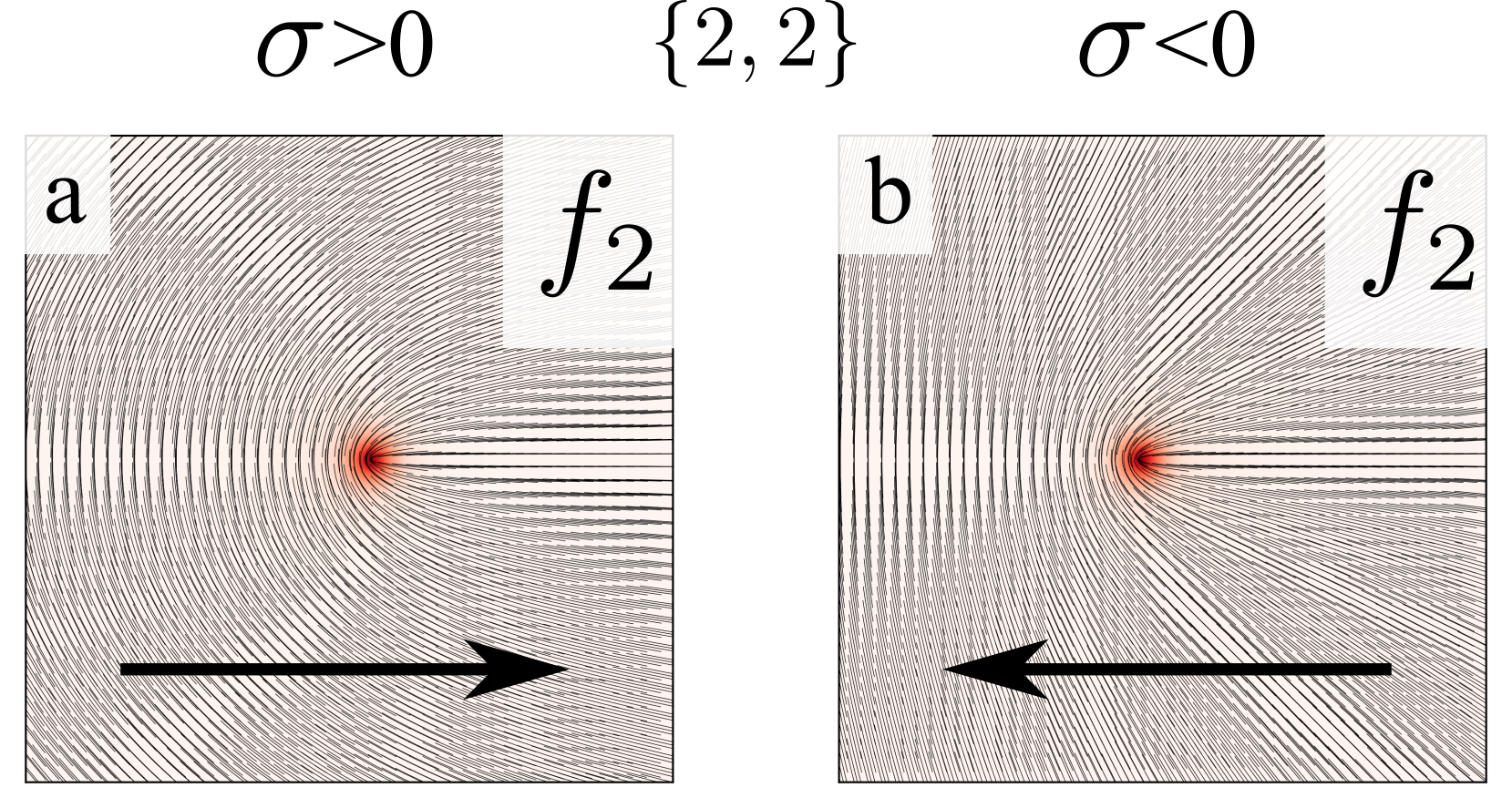}
	\caption{\label{fig:S5a} Behavior of an isolated $+1/2$ defect in a $\{2,2\}$ system simulated using our hydrodynamic model. (a) $\sigma>0$ we see a defect with azimuthal alignment of the director around the defect head that self propels toward its tail (black arrow). (b) $\sigma<0$ we see a defect with radial alignment of the director around the defect tail that self propels toward its head (black arrow). } 
\end{figure}

\begin{figure}[h]
	\centering
	\includegraphics[width=0.5\columnwidth]{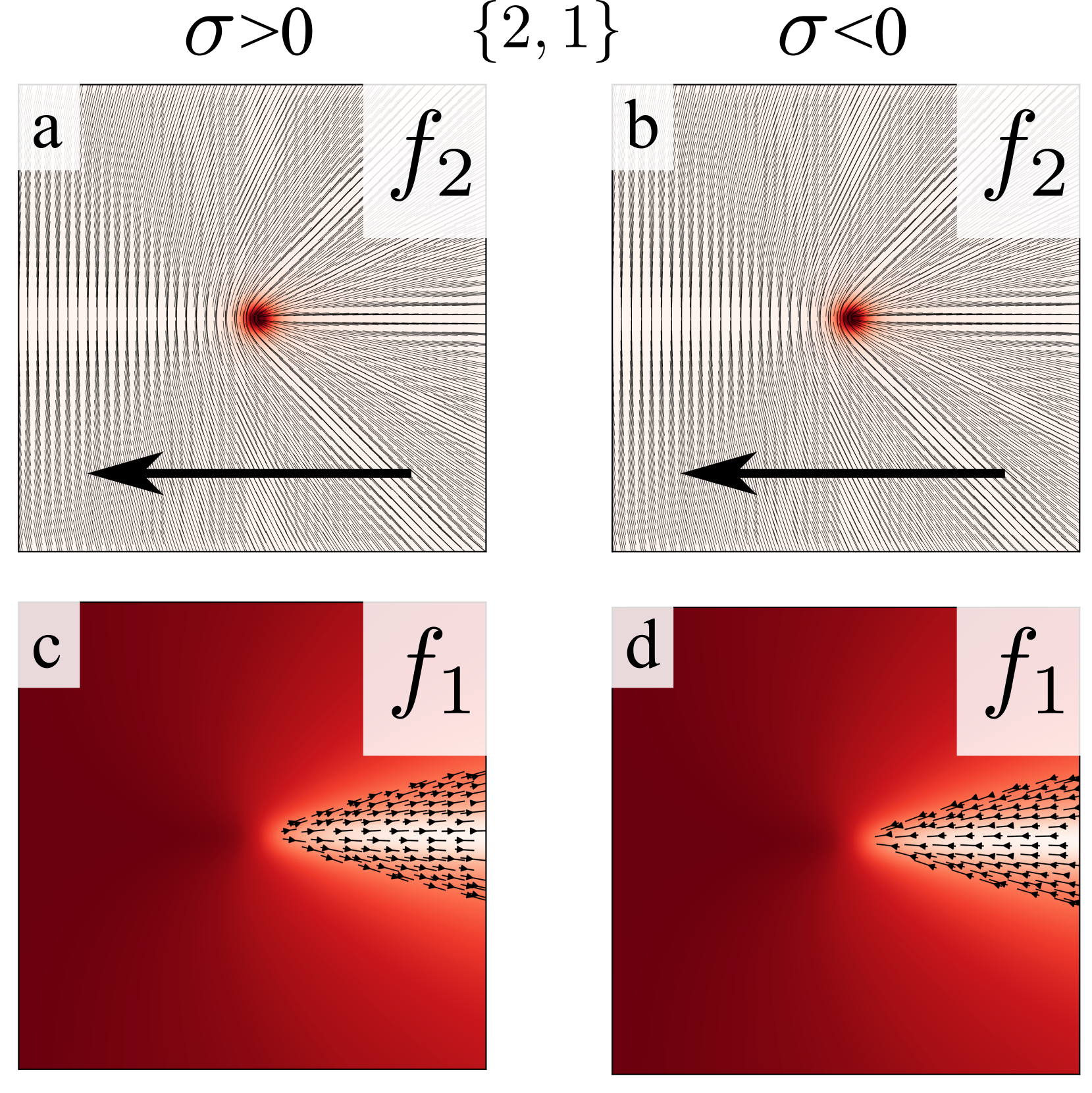}
	\caption{\label{fig:S5b} Behavior of an isolated $+1/2$ defect in a $\{2,1\}$ system simulated using our hydrodynamic model. Top row shows $f_2$, bottom row shows $f_1$. Left shows $\lambda_1>0$, right shows $\lambda_1<0$. (a)\&(b) The $f_2$ field shows a radially aligned director around the defect tail with self propulsion toward the head independent of the sign of $\lambda_1$. (c)\&(d) The $f_1$ field becomes ordered at the center of the defect. This leaves a trail of order in its wake as the defect moves. The $f_1$ field points away from the defect for $\lambda_1>0$ (c) and towards the defect for $\lambda_1<0$ (d).} 
\end{figure}

\FloatBarrier
\section{Derivation of hydrodynamic equations}
\subsection{Lattice model}
Consider $N$ lattice sites labeled by $i=1,2,...N$ at positions $\mathbf{R}_{i} \in \mathbb{R}^{2}$. At each lattice site $\mathbf{R}_{i}$ there is a spin $\theta_{i}$.

We associate the order of symmetry $\mathfrak{p}$ with the spins, such that two spins at sites $i$ and $j$ are considered aligned if $\theta_i = \theta_j+2\pi n/\mathfrak{p}$ for any integer $n$ where $\mathfrak{p}\in \mathbb{N}$. 

Each spin tries to align with its neighbors, however the interaction strength is modulated by the position of the neighbor relative to the orientation of the spin. We refer to this phenomenon as a vision cone. The vision cone has order of symmetry $\mathfrak{q} \in \mathbb{W}$. 
\newline 
\newline
The \emph{Stochastic Dynamics} of the spin is governed by
\begin{equation}\label{EQNR002}
	\begin{aligned}
		\dot{\theta}_{i} (t) = \eta_{i}(t) + J\sum^{N}_{j\neq i; \ j=1} K_{\kappa}(|\mathbf{s}|) \cdot  \left[1 + \sigma \cos \left( \mathfrak{q} \left( \theta_{i} - \alpha_{ij} (\mathbf{s}) \right) \right) \right] \cdot \sin \left( \mathfrak{p} \left( \theta_{j} - \theta_{i} \right) \right) 
	\end{aligned}
\end{equation}
where $J$ is an interaction constant, $\sigma$ is the strength of the non-reciprocal modulation and $\alpha_{ij}$ is defined as
\begin{equation}\label{EQNR003}
	\begin{aligned}
		\mathbf{s} = \mathbf{R}_{j} - \mathbf{R}_{i} \quad ; \quad s =  |\mathbf{s}|  \quad ; \quad  \alpha_{ij}(\mathbf{s}) = \textrm{arg}(\mathbf{s}).
	\end{aligned}
\end{equation}
We introduce the smooth interaction kernel $K_\kappa$
\begin{equation}\label{EQNR004}
	\begin{aligned}
		K_{\kappa}(|\mathbf{s}|) = \frac{1}{2\pi \kappa^{2}} \exp \left( - \frac{|\mathbf{s}|^{2}}{2 \kappa^{2}} \right) \quad \textrm{where} \quad \int d^{2} \mathbf{s} \ K_{\kappa}(|\mathbf{s}|) = 1
	\end{aligned}
\end{equation}

In the continuum derivation we use the normalized, untruncated Gaussian kernel, while the lattice simulations employ its discretized and truncated counterpart. Finally noise term $\eta_{i}(t)$ refers to \emph{Gaussian white noise} with 
\begin{equation}\label{EQNR005}
	\begin{aligned}
		\langle \eta_{i} (t) \rangle = 0 \quad ; \quad \langle \eta_{i}(t)\, \eta_{j} (t')\rangle = 2 D\, \delta_{ij} \,\delta(t-t')
	\end{aligned}
\end{equation}

\subsection{Coarse graining}

%Define the probability density of finding spin $i$ at position $\mathbf{r}$ and orientation $\theta$ to be
Introducing the microscopic orientational density associated with spin $i$, 
\begin{equation}\label{EQNR015}
	\begin{aligned}
		c_{i}(\mathbf{r},\theta,t) = \delta(\mathbf{r}-\mathbf{R}_{i}) \cdot \delta(\theta-\theta_{i}(t)).
	\end{aligned}
\end{equation}
Here $c_{i}$ is a microscopic delta-density for a particular realization of the stochastic dynamics. The coarse-grained one-particle orientational distribution $c(\mathbf{r},\theta,t)$ obtained by averaging the microscopic density over noise realizations and over the fixed lattice. Normalizing it locally such that
\begin{equation}
	\begin{aligned}
		\int^{2\pi}_{0} d\theta \ c(\mathbf{r},\theta,t) =1
	\end{aligned}
\end{equation}
%Thus the total density is given by
%\begin{equation}\label{EQNR016}
%\begin{aligned}
%c(\mathbf{r},\theta,t) & = \frac{1}{N} \sum^{N}_{i=1} \delta(\mathbf{r}-\mathbf{R}_{i}) \cdot %\delta(\theta-\theta_{i}(t)) \\
%& = \frac{1}{N} \sum^{N}_{i=1} c_{i} (\mathbf{r},\theta,t)
%\end{aligned}
%\end{equation}
%The probability density function will evolve according to 
Since the lattice sites remain fixed, the local number density is constant. We therefore normalize the coarse-grained orientational distribution by this density, such that $f_{0}=1$. The microscopic density evolves according to  
\begin{equation}\label{EQNR017}
	\begin{aligned}
		\partial_{t} c_{i} (\mathbf{r},\theta,t) = \delta (\mathbf{r} - \mathbf{R}_{i}) \cdot \partial_{t} \left[ \delta(\theta - \theta_{i}(t)) \right].
	\end{aligned}
\end{equation}
Applying Itô's lemma and averaging over the angular noise gives the evolution equation for the one-particle orientational distribution. We employ a mean-field closure by factorizing the two-particle distribution into a product of one-particle distributions, neglecting connected correlations between lattice sites. This leads to the Fokker–Planck equation,
%Using the Ito's lemma and averaging over the angular noise gives an equation for the one-particle distribution, we are able to re-write the above equation from our lattice model, which leads us to the Fokker Planck equation
\begin{multline}\label{EQNR040}
	\partial_{t} c(\mathbf{r},\theta,t) =
	- \partial_{\theta}\left[ \left( J \int d^{2}s \ d \theta ' K_{\kappa}(|\mathbf{s}|) \cdot \left[1 + \sigma \cos ( \mathfrak{q}(\theta -\alpha)) \right] \cdot \sin (\mathfrak{p} (\theta' - \theta)) \cdot c(\mathbf{r}',\theta',t)  \right) c(\mathbf{r},\theta,t) \right] \\+ D \partial^{2}_{\theta} c(\mathbf{r},\theta,t)
\end{multline}

where, $\mathbf{r}' = \mathbf{r} + \mathbf{s}$, $\alpha = \textrm{arg}(\mathbf{s})$. We next write the orientational distribution as a linear combination of Fourier modes, i.e.
\begin{equation}\label{EQNR047}
	\begin{aligned}
		f_{n} (\mathbf{r},t) & = \int d \theta \ \left[c(\mathbf{r},\theta,t) \cdot e^{\mathrm{i}n\theta}\right] \\
		c(\mathbf{r},\theta,t) & = \frac{1}{2\pi} \sum_{n} f_{n}(\mathbf{r},t) e^{-\mathrm{i}n\theta}
	\end{aligned}
\end{equation}

Since $c$ is real, $f_{-n} = f^{*}_{n}$; $f_{0}=1$. Here the $f_n$ are the complex amplitudes of each Fourier modes and can be written as
\begin{equation}
	f_n = A_n\exp(in\theta_n),
\end{equation}
Here $A_n$ is the magnitude of the Fourier mode $n$ and $\theta_n$ is a phase associated with each mode. 

This Fourier expansion separates $c(\mathbf{r},\theta,t)$ into contributions from each orientational order of symmetry. Orthogonality of the Fourier modes allows us to project  the dynamics onto each Fourier mode; the modes remain non-linearly coupled. 

We do this by substituting Eq.~\ref{EQNR047} into Eq.~\ref{EQNR040} and retaining only the lowest order terms to arrive at a dynamical equation for Fourier mode $n$ given by, 
\begin{equation}\label{eq:fn}
	\begin{aligned}
		\partial_{t} f_{n} = & - n^{2} D f_{n} + \frac{nJ}{4} \bigg[ f_{n-\mathfrak{p}} \left(2 + \kappa^{2} \Delta \right) f_{\mathfrak{p}}  - f_{n+\mathfrak{p}}  \left(2 + \kappa^{2} \Delta \right) f_{- \mathfrak{p}} \bigg] \\
		& + \frac{nJ\sigma'}{4} \bigg[ f_{n-\mathfrak{p}+\mathfrak{q}} \cdot \partial^{\mathfrak{q}}_{-} f_{\mathfrak{p}} - f_{n+\mathfrak{p}+\mathfrak{q}} \cdot \partial^{\mathfrak{q}}_{-} f_{-\mathfrak{p}} + f_{n-\mathfrak{p}-\mathfrak{q}} \cdot \partial^{\mathfrak{q}}_{+} f_{\mathfrak{p}} - f_{n+\mathfrak{p}-\mathfrak{q}} \cdot \partial^{\mathfrak{q}}_{+} f_{-\mathfrak{p}} \bigg]
	\end{aligned}
\end{equation}
Where we have introduced $\partial_\pm^\mathfrak{q} = (\partial_x \pm i\partial_y)^\mathfrak{q}$. We have also rescaled $\sigma$ by a factor of:
\begin{equation}
	\sigma' = \sigma \frac{\Gamma\left(\frac{\mathfrak{q}}{2} + 1 \right)}{2^{\frac{\mathfrak{q}}{2}}\mathfrak{q}!} \kappa^{\mathfrak{q}}.
\end{equation}
This indicates that the non-reciprocal contribution first appears at the \(q\)-th order in spatial gradients and therefore becomes progressively higher order in the long-wavelength expansion as \(q\) is increased.
%This indicates that the effect of non-reciprocity is reduced as $\mathfrak{q}$ is increased, i.e, the $\mathfrak{q}-$fold contribution first appears at the $\mathfrak{q}-$the spatial gradient order and is therefore progressively higher order in a long-wavelength expansion. 
%\ylann{there is a problem of units here, $\kappa$ has units of space. I assume one should have $\kappa / a$ instead, where $a$ is the lattice spacing ? \\ \dan{The new parameter $\sigma'$ is no longer dimensionless. The $k^q$ will cancel with the dimensions of the derivatives that are raised to the power of $q$. Also, there is no lattice spacing in this model.}
	%It is true that with $\kappa = 3/\sqrt{2}$, it is monotonically decreasing. \\ \dan{I am not sure I understand this comment, we don't give a value for kappa. The length you are quoting is the kernel width for the lattice model. Kappa is by definition smaller than the hydrodynamic length scales in our model. I guess this relationship gives you an idea of how the hydrodynamic length must scale with q.}
	%But actually, depending on $\kappa$, this function as a maximum. To be clarified.}

\subsection{Critical mode selection}

A spatially homogeneous solution to Eq.~\ref{eq:fn} is simply given by $f_n=0$ for all $n\neq0$. This corresponds to the globally disordered state, in which there is no-orientational order anywhere in the system. The $f_{0}$ mode describes the total probability of finding a spin with any orientation (at a certain location) and is thus proportional to the density and is set to be $1$ by normalization. Since our spins are fixed to a lattice, the density and hence $f_{0}$, is fixed to a constant.

We linearize about this homogeneous disordered state to identify which orientational mode first becomes unstable. Consider a spatially homogeneous perturbation with $|f_{n}|\ll 1$. Since there are no gradients due to it being spatially homogeneous, Eq.~\ref{eq:fn} to
\begin{equation}\label{eq:fn_dis}
	\partial_{t} f_{n} = - n^{2} D f_{n} + \frac{nJ}{2} \bigg[ f_{n-\mathfrak{p}} f_{\mathfrak{p}}  - f_{n+\mathfrak{p}}   f_{- \mathfrak{p}} \bigg].
\end{equation}

At linear order about the disordered state, the only orientational sector that receives an interaction-driven growth term is $n=\pm \mathfrak{p}$. Since out distributions are real, i.e., $f_{-\mathfrak{p}} = f^{*}_{\mathfrak{p}}$, these correspond to the same physical $\mathfrak{p}$-mode order parameter, and is sufficient to consider $n=\mathfrak{p}$ 
\begin{equation}\label{eq:fn_dis}
	\partial_{t} f_{\mathfrak{p}} = \mathfrak{p}\bigg[\frac{Jf_0}{2} - \mathfrak{p}D \bigg]f_{\mathfrak{p}}
\end{equation}
which gives the growth of the $f_{\mathfrak{p}}$ mode order 
\begin{equation}
	D<\frac{Jf_0}{2\mathfrak{p}} = D_{c}.
\end{equation}
Thus $f_{\mathfrak{p}}$ is the primary critical mode of the homogeneous order transition, i.e., $f_{\mathfrak{p}}$ is the only orientational mode that becomes linearly unstable at the homogeneous disordered-state transition.
% \ylann{$D_c = \frac{Jf_0}{2\gamma\mathfrak{p}}$. units ?}
% \ylann{Note for us : Here everything holds also for the mode $-\p$. Maybe we can simplify the discussion in the main text by only considering the positive modes.}

\subsubsection{Secondary mode selection}

We have established that the $f_{\mathfrak{p}}$ mode is the primary critical mode and spontaneously develops a finite amplitude below the transition. This does not imply that the remaining Fourier modes remain zero in the ordered phase. Once $f_{\mathfrak{p}} \neq 0$, (non-linear) mode coupling can generate additional harmonics, which are generally non-critical and slaved to this critical mode, $f_{\mathfrak{p}}$. For particular relations between $\mathfrak{p}$ and $\mathfrak{q}$, an additional mode may become anomalously slow and must be retained dynamically. 
A particularly important situation arises when $\mathfrak{p}=2\mathfrak{q}$. In this case, the homogeneous mode equation for $\mathfrak{q}$ contains a coupling proportional to $f_{\mathfrak{2q}} f_{-\mathfrak{q}}$. Since $f_{-\mathfrak{q}}=f^{*}_{\mathfrak{q}}$, the primary ordered mode couples the secondary mode directly to its complex conjugate, potentially reducing its relaxation rate. This distinguishes the $\mathfrak{p}=2\mathfrak{q}$ case from the other symmetry combinations and motivates retaining $f_{\mathfrak{q}}$ when its relaxation becomes sufficiently slow. This secondary mode is not independently critical at the disordered-state transition, but may become weakly damped within the ordered state.
We will come to this in more detail later.

\subsection{Solving for fast and slow hydrodynamic modes}

%The primary critical mode  $f_{\mathfrak{p}}$ becomes slow close to the ordering transition because its linear relaxation rate vanishes at criticality. Other modes need not be critical to be slow; conversely, modes whose relaxation rates remain finite at the transition are fast relative to $f_{\mathfrak{p}}$ and relax on a shorter timescale. This separation of timescales allows us to close up the hydrodynamic equations in terms of slow modes. 
Close to the ordering transition, the primary mode $f_{\mathfrak{p}}$ becomes critical, meaning that its linear relaxation rate vanishes at $D=D_{c}$. Consequently, it evolves on a long timescale and is a slow mode. More generally, we call a mode slow when its relaxation rate becomes small on the hydrodynamic timescale. Modes whose relaxation rates remain finite at the transition are fast modes and relax on a shorter timescale; to leading order, their time derivatives may therefore be set to zero and the modes expressed in terms of the retained slow fields.
Writing Eq.~\ref{eq:fn} for the primary critical mode gives:
\begin{equation}\label{eq:fp}
	\begin{aligned}
		\partial_{t} f_{\mathfrak{p}} = & - \mathfrak{p}^{2} D f_{\mathfrak{p}} + \frac{\mathfrak{p}J}{4} \bigg[ f_{0} \left(2 + \kappa^{2} \Delta \right) f_{\mathfrak{p}}  - f_{2\mathfrak{p}}  \left(2 + \kappa^{2} \Delta \right) f_{- \mathfrak{p}} \bigg] \\
		& + \frac{\mathfrak{p}J\sigma'}{4} \bigg[ f_{\mathfrak{q}} \cdot \partial^{\mathfrak{q}}_{-} f_{\mathfrak{p}} - f_{2\mathfrak{p}+\mathfrak{q}} \cdot \partial^{\mathfrak{q}}_{-} f_{-\mathfrak{p}} + f_{-\mathfrak{q}} \cdot \partial^{\mathfrak{q}}_{+} f_{\mathfrak{p}} - f_{2\mathfrak{p}-\mathfrak{q}} \cdot \partial^{\mathfrak{q}}_{+} f_{-\mathfrak{p}} \bigg]
	\end{aligned}
\end{equation}

This couples the primary critical mode of the system to a number of additional non-critical modes, $\big[f_{2\mathfrak{p}},f_{2\mathfrak{p}\pm\mathfrak{q}},f_{\pm\mathfrak{q}}\big]$. Whether these additional modes may be treated as fast or must also be retained as slow modes depends on their relaxation rates and scaling close to the transition. It is useful to first determine the scaling associated with the primary critical mode. 
% It is worth taking the time to consider how each of the terms scale close to the instability.
% . We define $\varepsilon^2 = \mathfrak{p}\bigg[\frac{Jf_0}{2} - \mathfrak{p}D \bigg]$, which represents the distance to the critical temperature. 
Writing Eq.~\ref{eq:fp} to linear order close to the instability we arrive at 
\begin{equation}
	\partial_t f_\mathfrak{p} = \varepsilon^2f_\mathfrak{p} + \frac{\mathfrak{p}Jf_0\kappa^2}{4}\Delta f_\mathfrak{p}
\end{equation}
where we have introduced the control parameter $\varepsilon^2 = \mathfrak{p}\bigg[\frac{Jf_0}{2} - \mathfrak{p}D \bigg]\ll1$. 
% \ylann{it has units, we don't want that. instead, I would suggest $\varepsilon^2 = (D_c - D)/D_c$, with $D_c = \frac{Jf_0}{2\p}$. }
Solving this in Fourier space allows us to determine that slow modes have spatial variations that vary with $\varepsilon$. Thus we can estimate how gradients vary close to instability, i.e. balancing the linear growth and diffusive terms the equation above gives the long-wavelength scaling, $\Delta\sim\varepsilon^2$ and $\partial_\pm^\mathfrak{q}\sim \varepsilon^\mathfrak{q}$.

We next consider the additional modes appearing in Eq.~\ref{eq:fp}. When such a mode retains a finite linear relaxation rate at the primary transition, it is fast relative to $f_{\mathfrak{p}}$. Its dynamics therefore relax on a shorter timescale than that of the slow mode, allowing us to set its time derivative zero to leading order. For example substituting $n=2\mathfrak{p}$ into Eq.~\ref{eq:fn}. 
%We next write the dynamical equations for each of the fast modes featured in Eq.~\ref{eq:fp}. For example, substituting $n=2\mathfrak{p}$ into Eq.~\ref{eq:fn} gives

\begin{equation}\label{eq:f2p}
	\begin{aligned}
		0 \simeq \partial_{t} f_{2\mathfrak{p}} = & - 4\mathfrak{p}^{2} D f_{2\mathfrak{p}} + \frac{\mathfrak{p}J}{2} \bigg[ f_{\mathfrak{p}} \left(2 + \kappa^{2} \Delta \right) f_{\mathfrak{p}}  - f_{3\mathfrak{p}}  \left(2 + \kappa^{2} \Delta \right) f_{- \mathfrak{p}} \bigg] \\
		& + \frac{\mathfrak{p}J\sigma'}{2} \bigg[ f_{\mathfrak{p}+\mathfrak{q}} \cdot \partial^{\mathfrak{q}}_{-} f_{\mathfrak{p}} - f_{3\mathfrak{p}+\mathfrak{q}} \cdot \partial^{\mathfrak{q}}_{-} f_{-\mathfrak{p}} + f_{\mathfrak{p}-\mathfrak{q}} \cdot \partial^{\mathfrak{q}}_{+} f_{\mathfrak{p}} - f_{3\mathfrak{p}-\mathfrak{q}} \cdot \partial^{\mathfrak{q}}_{+} f_{-\mathfrak{p}} \bigg]
	\end{aligned}
\end{equation}

The lowest-order forcing term then gives the scaling close to the instability as:
\begin{equation}
	f_{2\mathfrak{p}}\sim f_\mathfrak{p}^2.
\end{equation}
For the hierarchy of the higher harmonics generated by the primary mode, this extends inductively as
\begin{equation}
	f_{n\mathfrak{p}}\sim f_{\mathfrak{p}}^n.
\end{equation}
Since the distribution is real, $f_{-\mathfrak{p}}= f^{*}_{\mathfrak{p}}$, $f_{-\mathfrak{p}}$ has the same scaling as $f_{\mathfrak{p}}$. This allows us to estimate the scale of the first non-linear term in Eq.~\ref{eq:fp} as $f_{2\mathfrak{p}}f_{-\mathfrak{p}}\sim f_{\mathfrak{p}}^3$. Balancing this with the linear terms close to the instability allows us to estimate the scaling of $f_\mathfrak{p}$ close to the instability as:
\begin{equation}
	f_\mathfrak{p}\sim \varepsilon.
\end{equation}

These scalings allow us to write a closed set of hydrodynamic equations for the retained slow mode(s) to a known order in $\varepsilon$. 

The resulting expansion is controlled close to the ordering transition, where the primary mode is weakly ordered and spatial gradients are small. The scaling of additional non-primary modes must be determined separately when their relaxation rates become small.

\subsubsection{Dynamical equations when $\mathfrak{p} = \mathfrak{q}$.}

We do this now for $\mathfrak{p} = \mathfrak{q}$, all other combinations follow a similar methodology but can result in different terms becoming relevant. 

The slow mode equation becomes 
\begin{equation}\label{eq:fpq}
	\begin{aligned}
		\partial_{t} f_{\mathfrak{p}} = & - \mathfrak{p}^{2} D f_{\mathfrak{p}} + \frac{\mathfrak{p}J}{4} \bigg[ f_{0} \left(2 + \kappa^{2} \Delta \right) f_{\mathfrak{p}}  - f_{2\mathfrak{p}}  \left(2 + \kappa^{2} \Delta \right) f_{- \mathfrak{p}} \bigg] \\
		& + \frac{\mathfrak{p}J\sigma'}{4} \bigg[ f_{\mathfrak{p}} \cdot \partial^{\mathfrak{p}}_{-} f_{\mathfrak{p}} - f_{3\mathfrak{p}} \cdot \partial^{\mathfrak{p}}_{-} f_{-\mathfrak{p}} + f_{-\mathfrak{p}} \cdot \partial^{\mathfrak{p}}_{+} f_{\mathfrak{p}} - f_{\mathfrak{p}} \cdot \partial^{\mathfrak{p}}_{+} f_{-\mathfrak{p}} \bigg]
	\end{aligned}
\end{equation}

%Hence we only need the fast mode equations for $f_{2\mathfrak{p}}$ and $f_{3\mathfrak{p}}$. 
The additional modes $f_{2\mathfrak{p}}$ and $f_{3\mathfrak{p}}$ remain fast near the ordering transition. Setting their time derivatives to zero to leading order gives us the equations needed to write a closed set of hydrodynamic equations.

Solving Eq.~\ref{eq:f2p} to order $\varepsilon^3$ with $\mathfrak{p} = \mathfrak{q}$ gives us:
\begin{equation}\label{eq:f2p2}
	\begin{aligned}
		\frac{8\mathfrak{p} D}{J}f_{\mathfrak{2p}} &= 2f_{\mathfrak{p}}^2+ \sigma'f_{0} \cdot \partial^{\mathfrak{p}}_{+} f_{\mathfrak{p}}
	\end{aligned}
\end{equation}
Note here that the final term has order $\varepsilon^{\mathfrak{p}+1}$ so is negligible at higher $\mathfrak{p}$.

Similarly solving the $f_{3\mathfrak{p}}$ equation to order $\varepsilon^3$ gives us
\begin{equation}\label{eq:f3p}
	\begin{aligned}
		\frac{12\mathfrak{p} D}{J} f_{3\mathfrak{p}} = 2f_{2\mathfrak{p}}f_{\mathfrak{p}}  + \sigma'  f_{\mathfrak{p}} \cdot \partial^{\mathfrak{p}}_{+} f_{\mathfrak{p}}.
	\end{aligned}
\end{equation}

Inserting these back into Eq.~\ref{eq:fpq} and writing all terms up to order $\varepsilon^4$ gives us
\begin{equation}\label{eq:fpf}
	\begin{aligned}
		\partial_{t} f_{\mathfrak{p}} = & \frac{\mathfrak{p}}{2}[Jf_0- 2\mathfrak{p} D] f_{\mathfrak{p}} + \frac{\mathfrak{p}J}{4} f_{0} \kappa^{2} \Delta  f_{\mathfrak{p}}- \frac{J^2}{8D}|f_\mathfrak{p}|^2f_\mathfrak{p}  - \frac{J^2\sigma'f_0}{16D}  f_{- \mathfrak{p}}\cdot\partial_+^\mathfrak{p}f_\mathfrak{p}     \\
		& + \frac{\mathfrak{p}J\sigma'}{4} \bigg[ f_{\mathfrak{p}} \cdot \partial^{\mathfrak{p}}_{-} f_{\mathfrak{p}}  + f_{-\mathfrak{p}} \cdot \partial^{\mathfrak{p}}_{+} f_{\mathfrak{p}} - f_{\mathfrak{p}} \cdot \partial^{\mathfrak{p}}_{+} f_{-\mathfrak{p}} \bigg]
	\end{aligned}
\end{equation}
Note that since $f_{3\mathfrak{p}}\sim\varepsilon^3$ is coupled with terms $\partial^{\mathfrak{p}}_{-} f_{-\mathfrak{p}}\sim\varepsilon^{1+p}$ it does not appear in our final equation at this order. 

Introducing simplified parameters we can write the hydrodynamic equation for the $\mathfrak{p}=\mathfrak{q}$ case to be
\begin{equation}\label{eq:fpf1}
	\begin{aligned}
		\partial_{t} f_{\mathfrak{p}} = & A(B-|f_\mathfrak{p}|^2)f_{\mathfrak{p}} + K \Delta  f_{\mathfrak{p}}    + \lambda_1 \bigg[ f_{\mathfrak{p}} \cdot \partial^{\mathfrak{p}}_{-} f_{\mathfrak{p}}  - f_{\mathfrak{p}} \cdot \partial^{\mathfrak{p}}_{+} f_{-\mathfrak{p}} \bigg] +  \lambda_2 f_{- \mathfrak{p}}\cdot\partial_+^\mathfrak{p}f_\mathfrak{p} 
	\end{aligned}
\end{equation}
Here  $A$ and $B$ determine the transition between order and disorder. $K$ acts like an elastic constant penalizing gradients in the orientation of the spins. Finally $\lambda_1$ and $\lambda_2$ are active terms that describe the action of non-reciprocity. The active terms scale with $\varepsilon^{2+\mathfrak{p}}$, thus we expect them to become negligible at order $\varepsilon^4$ for $\mathfrak{p}>2$. 

Within the controlled weak-order expansion close to ordering transition these coefficients are related to the microscopic parameters by: 
\begin{center}
	\begin{tabular}{ c c }
		&A = $\frac{J^2}{8D}$ $\quad ; \quad$ B = $\frac{4D\mathfrak{p}}{J^2}(Jf_0-2\mathfrak{p}D)$\\  
		&K = $\frac{\mathfrak{p}J\kappa^2 f_0}{4}$ \\ 
		&$\lambda_1
		$ = $\frac{\mathfrak{p}J\sigma'}{4}$ \quad ; \quad $\lambda_2$ = $\frac{\mathfrak{p}J\sigma'}{4} - \frac{J^2\sigma'f_0}{16D}$\end{tabular}
\end{center}

These expressions give the microscopic near-onset values of the continuum coefficients and therefore also impose relations between them in this regime. However, the expansion is controlled only close to the ordering transition, where $f_{\mathfrak{p}}\sim\varepsilon$ and gradients are small. In the strongly ordered, low-noise regime considered in much of this work, these quantitative coefficient relations are not expected to remain controlled.

We therefore use Eq.~\ref{eq:fpf1} more generally as an effective continuum theory: its operator structure is retained, while $A, \ B, \ K, \ \lambda_1, \ \lambda_2$ are treated as independent effective parameters away from the near-onset regime.

For $\mathfrak{p}>2$, the leading non-reciprocal terms lie beyond order $\varepsilon^{4}$. We nevertheless retain these symmetry-allowed operators to describe the leading non-reciprocal effects at higher symmetries, regarding the resulting equation as an effective continuum model beyond the strict fourth-order truncation.

\subsubsection{Dynamical equations when $ \mathfrak{q}=2\mathfrak{p}$.}

In this case the dynamical equation for the slow mode is given by:
\begin{equation}\label{eq:fp2q1}
	\begin{aligned}
		\partial_{t} f_{\mathfrak{p}} = & - \mathfrak{p}^{2} D f_{\mathfrak{p}} + \frac{\mathfrak{p}J}{4} \bigg[ f_{0} \left(2 + \kappa^{2} \Delta \right) f_{\mathfrak{p}}  - f_{2\mathfrak{p}}  \left(2 + \kappa^{2} \Delta \right) f_{- \mathfrak{p}} \bigg] \\
		& + \frac{\mathfrak{p}J\sigma'}{4} \bigg[ f_{2\mathfrak{p}} \cdot \partial^{2\mathfrak{p}}_{-} f_{\mathfrak{p}} - f_{4\mathfrak{p}} \cdot \partial^{2\mathfrak{p}}_{-} f_{-\mathfrak{p}} + f_{-2\mathfrak{p}} \cdot \partial^{2\mathfrak{p}}_{+} f_{\mathfrak{p}} - f_{0} \cdot \partial^{2\mathfrak{p}}_{+} f_{-\mathfrak{p}} \bigg]
	\end{aligned}
\end{equation}

Clearly the first terms are identical to the $\mathfrak{p}=\mathfrak{q}$ case and will result in similar terms. 

The remaining non-reciprocal terms scale with higher orders of $\varepsilon$ except the final term. Thus we only need to solve one additional fast mode equation which gives:
\begin{equation}\label{eq:f2p3}
	\begin{aligned}
		\frac{4\mathfrak{p} D}{J}f_{2\mathfrak{p}} &= f_{\mathfrak{p}}^2
	\end{aligned}
\end{equation}
to order $\varepsilon^3$.

We then arrive at the hydrodynamic equation:
\begin{equation}\label{eq:fpf2}
	\begin{aligned}
		\partial_{t} f_{\mathfrak{p}} = & \frac{\mathfrak{p}}{2}[Jf_0- 2\mathfrak{p} D] f_{\mathfrak{p}} + \frac{\mathfrak{p}J}{4} f_{0} \kappa^{2} \Delta  f_{\mathfrak{p}}- \frac{J^2}{8D}|f_\mathfrak{p}|^2f_\mathfrak{p}      - \frac{\mathfrak{p}J\sigma'f_0}{4}   \partial^{2\mathfrak{p}}_{+} f_{-\mathfrak{p}} 
	\end{aligned}
\end{equation}

Introducing simplified coefficients we arrive at:
\begin{equation}\label{eq:fpf}
	\begin{aligned}
		\partial_{t} f_{\mathfrak{p}} = & A(B-|f_\mathfrak{p}|^2)f_{\mathfrak{p}} + K \Delta  f_{\mathfrak{p}} -  \lambda_1 \partial_+^{2\mathfrak{p}}f_{-\mathfrak{p}  }
	\end{aligned}
\end{equation}

The coefficients are given by:
\begin{center}
	\begin{tabular}{ c c }
		&A = $\frac{J^2}{8D}$ $\quad ; \quad$ B = $\frac{4D\mathfrak{p}}{J^2}(Jf_0-2\mathfrak{p}D)$\\  
		&K = $\frac{\mathfrak{p}J\kappa^2 f_0}{4}$ \\ 
		&$\lambda_1$ = $\frac{\mathfrak{p}J\sigma'f_0}{4}$ \\  
	\end{tabular}
\end{center}
Similar to the $\mathfrak{p}=\mathfrak{q}$ case, away from onset, we retain the operator structure of Eq.~\ref{eq:fpf} but treat \(A\), \(B\), \(K\), and \(\lambda_1\) as independent effective continuum parameters.

\subsubsection{Dynamical equations when $\mathfrak{p} = 2\mathfrak{q}$.}

%We see from our lattice simulations that the $\mathfrak{p} = 2\mathfrak{q}$ appears to have two slow modes, $f_\mathfrak{p}$ and $f_\mathfrak{q}$. 
In the $\mathfrak{p}=2\mathfrak{q}$ case, $f_{\mathfrak{2q}}$ is the primary critical mode, while $\mathfrak{q}$ represents an additional orientational mode coupled to the primary order. From our lattice simulations, we observe that this secondary mode can play an important role in the dynamics, particularly around topological defects. We therefore retain both $f_{\mathfrak{2q}}$ and $f_{\mathfrak{q}}$ in our coupled hydrodynamic description. The equation for the primary mode is given by

The $f_\mathfrak{p}$ equation is written as follows. 
\begin{equation}\label{eq:f2q}
	\begin{aligned}
		\partial_{t} f_{2\mathfrak{q}} = & - 4\mathfrak{q}^{2} D f_{2\mathfrak{q}} + \frac{\mathfrak{q}J}{2} \bigg[ f_{0} \left(2 + \kappa^{2} \Delta \right) f_{2\mathfrak{q}}  - f_{4\mathfrak{q}}  \left(2 + \kappa^{2} \Delta \right) f_{- 2\mathfrak{q}} \bigg] \\
		& + \frac{\mathfrak{q}J\sigma'}{2} \bigg[ f_{\mathfrak{q}} \cdot \partial^{\mathfrak{q}}_{-} f_{2\mathfrak{q}} - f_{5\mathfrak{q}} \cdot \partial^{\mathfrak{q}}_{-} f_{-2\mathfrak{q}} + f_{-\mathfrak{q}} \cdot \partial^{\mathfrak{q}}_{+} f_{2\mathfrak{q}} - f_{3\mathfrak{q}} \cdot \partial^{\mathfrak{q}}_{+} f_{-2\mathfrak{q}} \bigg]
	\end{aligned}
\end{equation}

By observing the linear terms we can see that this is our critical mode, with similar scaling as in the $\mathfrak{p}=\mathfrak{q}$ case. As above, we can write $f_{2n\mathfrak{q}}\sim f_{2\mathfrak{q}}^n$. This allows us to estimate the scale of fast modes that are even in $\mathfrak{q}$.

Again considering to order $\varepsilon^3$ we obtain 
\begin{equation}
	\frac{16\mathfrak{q}D}{J}f_{4\mathfrak{q}} = 2f_{2\mathfrak{q}}^2 + \sigma'f_\mathfrak{q}\partial_+^\mathfrak{q}f_{2\mathfrak{q}}
\end{equation}
Introducing this into the above equation again lets us determine that $f_{2\mathfrak{q}}\sim \varepsilon$. Note, once again the scaling of the final term depends on the value of $\mathfrak{q}$ so will become higher order for increased $\mathfrak{q}$. It also depends on the scaling of $f_\mathfrak{q}$ that we do not yet know.

%Using a similar inductive argument, we can write a scaling law for the odd in $\mathfrak{q}$ fast modes which follows $f_{(2n+1)\mathfrak{q}}\sim f_\mathfrak{q}f_{2\mathfrak{q}}^n$. Closing this would require we know the scaling of the $f_\mathfrak{q}$ mode, which from our lattice model we assume to be slow. 
Similarly, for the homogeneous nonlinear couplings, the odd harmonics follow the scaling $f_{(2n+1)\mathfrak{q}}\sim f_{\mathfrak{q}} f_{2\mathfrak{q}}^{n}$. However, in spatially inhomogeneous systems, the non-reciprocal terms can also generate these modes through gradients of $f_{2\mathfrak{q}}$. The scaling of $\mathfrak{q}$ must therefore be considered separately, while the higher harmonics may be eliminated in terms of the retained fields. 

Considering to order $\varepsilon^3$ we obtain:
\begin{equation}
	\frac{12\mathfrak{q}D}{J}f_{3\mathfrak{q}} = 2f_{2\mathfrak{q}}f_{\mathfrak{q}} + \sigma'\big[f_{2\mathfrak{q}}\partial_-^\mathfrak{q}f_{2\mathfrak{q}} + f_0\partial_+^\mathfrak{q}f_{2\mathfrak{q}}\big]
\end{equation}

Substituting these into Eq.~\ref{eq:f2q} allows us to write
\begin{equation}\label{eq:f2q2}
	\begin{aligned}
		\partial_{t} f_{2\mathfrak{q}} = &\mathfrak{q}\bigg[Jf_0 - 4\mathfrak{q} D\bigg] f_{2\mathfrak{q}} + \frac{\mathfrak{q}J\kappa^{2}f_0}{2}   \Delta  f_{2\mathfrak{q}}  -\frac{J^2}{8D} |f_{ 2\mathfrak{q}}|^2   f_{ 2\mathfrak{q}} -\frac{J^2\sigma'}{16D} f_{-2\mathfrak{q}}\cdot f_{\mathfrak{q}}\cdot\partial_+^\mathfrak{q}f_{2\mathfrak{q}}\\
		& + \frac{\mathfrak{q}J\sigma'}{2} \bigg[ f_{\mathfrak{q}} \cdot \partial^{\mathfrak{q}}_{-} f_{2\mathfrak{q}}  + f_{-\mathfrak{q}} \cdot \partial^{\mathfrak{q}}_{+} f_{2\mathfrak{q}}\bigg]\\
		&- \frac{J^2\sigma'}{24D} \bigg[ 2f_{\mathfrak{q}}\cdot f_{2\mathfrak{q}} \cdot \partial^{\mathfrak{q}}_{+} f_{-2\mathfrak{q}}  + \sigma'f_0\partial^{\mathfrak{q}}_{+} f_{2\mathfrak{q}}\cdot \partial^{\mathfrak{q}}_{+} f_{-2\mathfrak{q}}\bigg]
	\end{aligned}
\end{equation}
to order $\varepsilon^4$; note here we have assumed the scaling $f_\mathfrak{q}\leq\varepsilon$. 

For $\mathfrak{q}$, the terms retained in Eq.~.\ref{eq:f2q} contribute up to order $\varepsilon^{4}$, assuming $f_{\mathfrak{q}}=O(\varepsilon)$. This scaling corresponds to retaining the secondary field as an independent dynamical variable. For higher values of $\mathfrak{q}$, some non-reciprocal terms enter at higher orders in the gradient expansion, but are retained as the leading contributions associated with their respective operators.

We next consider the equation for the $f_\mathfrak{q}$ mode which has dynamical equation:
\begin{equation}
	\begin{aligned}
		\partial_{t} f_{\mathfrak{q}} = & - \mathfrak{q}^{2} D f_{\mathfrak{q}} + \frac{\mathfrak{q}J}{4} \bigg[ f_{-\mathfrak{q}} \left(2 + \kappa^{2} \Delta \right) f_{2\mathfrak{q}}  - f_{3\mathfrak{q}}  \left(2 + \kappa^{2} \Delta \right) f_{- 2\mathfrak{q}} \bigg] \\
		& + \frac{\mathfrak{q}J\sigma'}{4} \bigg[ f_{0} \cdot \partial^{\mathfrak{q}}_{-} f_{2\mathfrak{q}} - f_{4\mathfrak{q}} \cdot \partial^{\mathfrak{q}}_{-} f_{-2\mathfrak{q}} + f_{-2\mathfrak{q}} \cdot \partial^{\mathfrak{q}}_{+} f_{2\mathfrak{q}} - f_{2\mathfrak{q}} \cdot \partial^{\mathfrak{q}}_{+} f_{-2\mathfrak{q}} \bigg]
	\end{aligned}
\end{equation}

Considering the linear terms in $f_{\pm\mathfrak{q}}$ of this equation gives:
\begin{equation}
	\begin{aligned}
		\partial_{t} f_{\mathfrak{q}} = - \mathfrak{q}^{2} D f_{\mathfrak{q}} + \frac{\mathfrak{q}J}{2} \cdot f_{-\mathfrak{q}} f_{2\mathfrak{q}}.
	\end{aligned}
\end{equation}
%The second term couples the growth of the $f_\mathfrak{q}$ mode to the amplitude of the critical $f_{2\mathfrak{q}}$ mode, this can lead to a secondary slow mode. When the critical mode $f_{2\mathfrak{q}}$ becomes ordered, it reduces the linear decay rate of the $f_\mathfrak{q}$ mode. This results in an increased relaxation time leading to slow mode behavior. In the regime where this decay is sufficiently slow, $f_{\mathfrak{q}}$ must be retained dynamically.
The second term couples the $f_{\mathfrak{q}}$ mode to the amplitude of the primary critical mode $f_{2\mathfrak{q}}$. Since $f_{-\mathfrak{q}}=f^{*}_{\mathfrak{q}}$, this coupling distinguishes polar fluctuations parallel and perpendicular to the ordered $f_{2\mathfrak{q}}$ field. In particular, the relaxation rate of one polar component is reduced as the primary mode becomes ordered. However, this coupling does not necessarily make $f_{\mathfrak{q}}$ independently critical. In the regime where the relaxation of $f_{\mathfrak{q}}$ becomes sufficiently slow compared with the dynamics of interest, it is necessary to retain $f_{\mathfrak{q}}$ as an independent dynamical field.

Once again we substitute for the fast modes and neglect higher order terms to arrive at:
\begin{equation}
	\begin{aligned}
		\partial_{t} f_{\mathfrak{q}} = & - \mathfrak{q}^{2} D f_{\mathfrak{q}} + \frac{\mathfrak{q}J}{4} \bigg[ f_{-\mathfrak{q}} \left(2 + \kappa^{2} \Delta \right) f_{2\mathfrak{q}}\bigg]  - \frac{J^2}{12D} |f_{2\mathfrak{q}}|^2f_{\mathfrak{q}}  - \frac{J^2\sigma'}{24D}|f_{2\mathfrak{q}}|^2\partial_-^\mathfrak{q}f_{2\mathfrak{q}} - \frac{J^2\sigma'f_0}{24D} f_{- 2\mathfrak{q}}\cdot \partial_+^\mathfrak{q}f_{2\mathfrak{q}}   \\
		& + \frac{\mathfrak{q}J\sigma'}{4} \bigg[ f_{0} \cdot \partial^{\mathfrak{q}}_{-} f_{2\mathfrak{q}} + f_{-2\mathfrak{q}} \cdot \partial^{\mathfrak{q}}_{+} f_{2\mathfrak{q}} - f_{2\mathfrak{q}} \cdot \partial^{\mathfrak{q}}_{+} f_{-2\mathfrak{q}} \bigg]\\
		& -\frac{J^2\sigma'}{32D}  f_{2\mathfrak{q}}^2 \cdot \partial^{\mathfrak{q}}_{-} f_{-2\mathfrak{q}}
	\end{aligned}
\end{equation}

Gathering all the terms and introducing simplified coefficients gives the following coupled equations: 
\begin{equation}\label{eq:f2_21}
	\begin{aligned}
		\partial_{t} f_{2\mathfrak{q}} = & A(B-|f_{ 2\mathfrak{q}}|^2)f_{2\mathfrak{q}} + K   \Delta  f_{2\mathfrak{q}} \\
		& + \lambda_{1} \left[ f_{\mathfrak{q}} \partial^{\mathfrak{q}}_{-} f_{2\mathfrak{q}} + f_{-\mathfrak{q}} \partial^{\mathfrak{q}}_{+} f_{2\mathfrak{q}}\right] \\ 
		& - \lambda_{2} \left( \partial^{\mathfrak{q}}_{+} f_{2\mathfrak{q}}\right) \left( \partial^{\mathfrak{q}}_{+} f_{-2\mathfrak{q}} \right) \\
		& - \lambda_{3} f_{-2\mathfrak{q}} f_{\mathfrak{q}} \partial^{\mathfrak{q}}_{+} f_{2\mathfrak{q}} \\
		& - \lambda_{4} f_{\mathfrak{q}} f_{2\mathfrak{q}} \partial^{\mathfrak{q}}_{+} f_{-2\mathfrak{q}} 
	\end{aligned}
\end{equation}
\begin{equation}\label{eq:f1_21}
	\begin{aligned}
		\partial_{t} f_{\mathfrak{q}} = & C_{1}f_{2\mathfrak{q}}\cdot f_{-\mathfrak{q}}- C_{2}f_{\mathfrak{q}} - C_{3}|f_{2\mathfrak{q}}|^2f_{\mathfrak{q}}  + \frac{K}{2} f_{-\mathfrak{q}}\Delta f_{2\mathfrak{q}}\\
		& + \frac{\lambda_{1}}{2} \left[ \partial^{\mathfrak{q}}_{-} f_{2\mathfrak{q}} + f_{-2\mathfrak{q}}\partial^{\mathfrak{q}}_{+} f_{2\mathfrak{q}} - f_{2\mathfrak{q}} \partial^{\mathfrak{q}}_{+} f_{-2\mathfrak{q}} \right] \\
		& - \lambda_{5} |f_{2\mathfrak{q}}|^{2} \partial^{\mathfrak{q}}_{-} f_{2\mathfrak{q}} \\
		& - \lambda_{6} f_{-2\mathfrak{q}}\partial^{\mathfrak{q}}_{+} f_{2\mathfrak{q}} \\
		& - \lambda_{7} f^{2}_{2\mathfrak{q}} \partial^{\mathfrak{q}}_{-} f_{-2\mathfrak{q}} 
	\end{aligned}    
\end{equation}

The coefficients are given by:
\begin{center}
	\begin{tabular}{ c c }
		&A = $\frac{J^2}{8D}$ $\quad ; \quad$ B = $\frac{8D\mathfrak{q}}{J^2}(Jf_0-4\mathfrak{q}D)$\\  
		&$C_{1}$  = $\frac{\mathfrak{q}J}{2}$  $\quad ; \quad $ $C_{2}$ = $\mathfrak{q}^{2}D$ $\quad ; \quad$ $C_{3}$ = $\frac{J^{2}}{12D}$ \\
		& K = $\frac{\mathfrak{q}J\kappa^2f_0}{2}$ \\ 
		& $\lambda_{1}=\frac{\mathfrak{q}J\sigma'}{2}$ $\quad ; \quad$ $\lambda_{2} = \frac{J^{2}f_{0}(\sigma')^2}{24D}$ \\
		& $\lambda_{3} = \frac{J^{2}\sigma'}{16D}$ $\quad ; \quad$ $\lambda_{4} = \frac{J^{2}\sigma'}{12D}$ $\quad ; \quad$ $\lambda_{5} = \frac{J^{2}\sigma'}{24D}$ \\
		& $\lambda_{6} \frac{J^{2} f_{0}\sigma'}{24D}$ $\quad ; \quad$ $\lambda_{7} = \frac{J^{2}\sigma'}{32D}$
	\end{tabular}
\end{center}

%These coefficient relations follow from the controlled near-onset expansion; away from onset, we retain \ref{eq:f2_21};\ref{eq:f1_21} as effective continuum equations and allow their coefficients to be treated as independent effective parameters.

The microscopic reduction generates seven effective gradient coefficients, whose bare values are related by the underlying interaction. We allow these coefficients to vary independently in the continuum description, while retaining the common coefficient $\lambda_1$ for the directly generated polar–nematic coupling in both field equations. The relative factor of two between these contributions follows from the Fourier-mode hierarchy.

The coefficient $\lambda_2$, associated with the quadratic-gradient contribution to the nematic equation, is generated through the elimination of higher harmonics and enters with an explicit negative sign. Near the primary nematic transition, when the polar mode retains a finite relaxation rate, the leading polar response scales as $f_1=O(\epsilon^2)$ for $f_2=O(\epsilon)$ and $\nabla=\mathcal{O}(\varepsilon)$. Consequently, the $\lambda_1$ and $\lambda_2$ sectors provide the leading nonlinear gradient contributions to the nematic dynamics at order $\mathcal{O}(\varepsilon^4)$, while the remaining gradient couplings contribute to the effective nematic equation at higher order.

We therefore retain $\lambda_1$ and $\lambda_2$ to obtain the minimal coupled continuum theory presented in Eq. (6) of the main text, setting the additional gradient coefficients to zero. This is a leading-order truncation of the near-onset nematic dynamics, rather than a complete truncation of the coupled two-field equations or a demonstration that the omitted operators are irrelevant in the ordered regime.

%\begin{center}
%\begin{tabular}{ c c }
%&A = $\frac{J^2}{8D}$ $\quad ; \quad$ B = $\frac{8D\mathfrak{q}}{J^2}(Jf_0-4\mathfrak{q}D)$\\  
%&$C_{1}$  = $\frac{\mathfrak{q}J}{2}$  $\quad ; \quad $ $C_{2}$ = $\mathfrak{q}^{2}D$ $\quad ; \quad$ $C_{3}$ = $\frac{J^{2}}{12D}$ \\
%& K = $\frac{\mathfrak{q}J\kappa^2f_0}{2}$ \\ 
%& $\lambda_1$ = $\frac{J^2\sigma'}{4D}$ $\quad ; \quad$ $\lambda_2$ = $\frac{\mathfrak{q}J\sigma'}{2}$  \\
%&$\lambda_3$ = -$\frac{J^2\sigma'^2f_0}{24D}$  $\quad ; \quad$ $\lambda_4$ = $\frac{J^2\sigma'f_0}{24D}$  
%\end{tabular}
%\end{center}

\section{Linear Stability analysis of the ordered state}

The defect self propulsion we observe in the $\{2,2\}$ and $\{2,1\}$ bears a resemblence to that of active nematics. In such systems, active stresses cause the ordered state to become unstable to perturbations that lead to defect proliferation and chaos. We examine the linear stability of the homogeneous ordered states to determine whether analogous instabilities arise in our continuum models.

%We demonstrate here that the same process is not possible in the systems studied here.   

\subsection{$\mathfrak{p}=2$, $\mathfrak{q}=2$}

We consider a uniform ordered nematic
\begin{equation}
	\begin{aligned}
		f_{2} = S_{0}e^{2\mathrm{i}\theta} \textrm{ where } S_{0}^{2}=B
	\end{aligned}
\end{equation}
with the reference director along the $x-$axis, and introduce a small orientational perturbation $\theta=\delta \theta$. To linear order,
\begin{equation}
	\begin{aligned}
		f_{\pm2} \simeq S_{0}
		\left( 1 \pm 2 \mathrm{i} \delta \theta \right)
	\end{aligned}
\end{equation}
For pure bend and pure splay perturbations the wave-vector lies respectively along the $x-$ and $y-$directions. In these two sectors the mixed amplitude-orientation derivative terms vanish, and the linear orientational equation becomes
\begin{equation}\label{EQNR066}
	\begin{aligned}
		\partial_{t} \delta\theta = \left[ K + S_{0} \left( \lambda_{2}+2\lambda_{1} \right) \right] \partial_{x} ^{2}\delta\theta + \left[ K - S_{0} \left( \lambda_{2} +2\lambda_{1} \right) \right] \partial_{y}^{2}\delta\theta.
	\end{aligned}
\end{equation}
% For a perturbation with wave vector $k$, the linearized equation above can only give a dispersion relation of the form $\omega\sim k^2$. This it is impossible for a finite range of unstable wave vectors, \ylann{I don't understand - Yuv : I have tried to write a similar statement in the paragraph below, let me know if that makes it clearer. } implying there is no route to well behaved chaotic states such as those observed in typical active nematics. 
For a Fourier perturbation with wave vector $k$, the linearized theory gives a dispersion relation of the form $\omega\sim -k^{2}$. Thus, at this order in gradients, there is no intrinsic finite-wavelength selection, preventing the mechanism giving rise to sustained chaotic states as observed in active nematics. 
%If either effective stiffness becomes negative, the instability instead grows toward short wavelengths, where higher-gradient terms or microscopic effects are required to regularize the dynamics.

For the reference director $\mathbf{n}=(1,0)$,
\begin{equation}
	\begin{aligned}
		\nabla\cdot\mathbf{n} &\simeq \partial_{y}\delta\theta \equiv \textrm{Splay} \\
		\left( \nabla\times\mathbf{n} \right) &\simeq \partial_{x}\delta\theta \equiv \textrm{Bend}
	\end{aligned}
\end{equation}
Hence the effective coefficients controlling pure bend and pure splay perturbations are
\begin{equation}\label{EQNR067}
	\begin{aligned}
		K_{\textrm{bend}} &= K + S_{0} \left( \lambda_{2}+2\lambda_{1}\right) \\
		K_{\textrm{splay}} &= K - S_{0} \left( \lambda_{2}+2\lambda_{1} \right).
	\end{aligned}
\end{equation}
For Fourier perturbations,
\begin{equation}
	\begin{aligned}
		\omega_{\textrm{bend}} &= -K_{\textrm{bend}}k^{2} \\
		\omega_{\textrm{splay}} &= -K_{\textrm{splay}}k^{2}
	\end{aligned}
\end{equation}

Therefore, provided both coefficients remain positive,
\begin{equation}
	\begin{aligned}
		\lambda_{2}+2\lambda_{1}>0 & \quad \Rightarrow \quad K_{\textrm{bend}}>K_{\textrm{splay}} \quad \Rightarrow \quad \textrm{bend perturbations decay faster than splay perturbations} \\
		\lambda_{2}+2\lambda_{1}<0 & \quad \Rightarrow \quad K_{\textrm{splay}}>K_{\textrm{bend}} \quad \Rightarrow \quad \textrm{splay perturbations decay faster than bend perturbations}
	\end{aligned}
\end{equation}

\subsection{$\mathfrak{p}=2$, $\mathfrak{q}=1$}
As in the $\mathfrak{p}=2;\mathfrak{q}=2$ model, it is useful to separately determine the response of a homogeneous ordered nematic to pure bend and pure splay perturbations. The structure is, however, qualitatively different because the non-reciprocal terms first generate a response in the secondary polar field. We consider
\begin{equation}
	\begin{aligned}
		f_{2} = S_{0}e^{2\mathrm{i}\delta\theta} \simeq S_{0}( 1 + 2\mathrm{i}\,\delta\theta) \quad ; \quad S_{0}^{2}=B \quad \textrm{and } f_{1} = P_{x}+\mathrm{i}P_{y}
	\end{aligned}
\end{equation}
The reference director is chosen along the $x-$axis, so, as before
\begin{equation}
	\begin{aligned}
		\partial_{y}\delta\theta  \equiv \textrm{Splay}  \quad ;  \quad \partial_{x}\delta\theta &\equiv \textrm{Bend}
	\end{aligned}
\end{equation}
At strict linear order about the homogeneous state $f_{1}=0$, all non-reciprocal terms in the $f_{2}$ equation are at least quadratic in the perturbations. The nematic orientation therefore obeys
\begin{equation}\label{EQNR088}
	\begin{aligned}
		\partial_t\delta\theta = K\Delta\delta\theta
	\end{aligned}
\end{equation}

Thus, like in the $\mathfrak{p}=2;\mathfrak{q}=2$ system, there is no route to sustained chaotic behavior, i.e., at this order, the homogeneous nematics orientation has only diffusive $k^{2}$ decay and no intrinsic finite-wavelength instability. However, unlike the $\mathfrak{p}=2;\mathfrak{q}=2$ model, there is no direct linear splitting of the nematic relaxation into separate $K_{\textrm{bend}}$ and $K_{\textrm{splay}}$ coefficients about the state $f_{1}=0$. The bend-splay distinction instead appears in the polar field generated by these deformations. Linearizing Eq.~\ref{eq:f1_21} gives 
\begin{equation}\label{EQNR089}
	\begin{aligned}
		\partial_{t}P_{x} &= - \left[ C_{2} + C_{3}S_{0}^{2} - C_{1}S_{0} \right]P_{x} + \left[ \lambda_{1}S_{0} \left( f_{0}-2S_{0} \right) + 2\lambda_{6}S_{0}^{2} + 2(\lambda_{7} -\lambda_{5})S_{0}^{3} \right] \partial_{y}\delta\theta \\
		\partial_{t}P_{y} &= - \left[ C_{2} + C_{3}S_{0}^{2} + C_{1}S_{0} \right]P_{y} + \left[ \lambda_{1}S_{0} \left( f_{0}+2S_{0} \right) - 2\lambda_{6}S_{0}^{2} + 2(\lambda_{7}-\lambda_{5})S_{0}^{3} \right] \partial_{x}\delta\theta
	\end{aligned}
\end{equation}
Hence,
\begin{equation}
	\begin{aligned}
		\textrm{Splay} \ \Rightarrow \  P_{x} \quad ; \quad \textrm{Bend} \ \Rightarrow\ P_{y}
	\end{aligned}
\end{equation}
Thus, a splay deformation generates a polar field parallel to the background nematic director, while a bend deformation generates a polar field perpendicular to the background nematic director. The sign of the corresponding source coefficient determines the polarity of that response.
\newline
\newline
The corresponding polar relaxation rates are
\begin{equation}
	\begin{aligned}
		\gamma_{\parallel} &=C_{2} + C_{3}S_{0}^{2} - C_{1}S_{0} \\
		\gamma_{\perp} &= C_{2} + C_{3} S_{0}^{2} + C_{1}S_{0}
	\end{aligned}
\end{equation}
For $C_{1}>0$ 
\begin{equation}
	\begin{aligned}
		\gamma_{\parallel} < \gamma_{\perp}
	\end{aligned}
\end{equation}
Therefore the polar response generated by a splay deformation is longer lived than the polar response generated by a bend deformation. In the $\mathfrak{p}=2;\mathfrak{q}=1$ model the natural distinction between bulk bend and splay is therefore not a direct difference between two nematic elastic relaxation coefficients, but rather a distinction between the secondary polar channels which the two distortions excite. This distinction becomes particularly strong close to the secondary-mode regime, where 
\begin{equation}
	\begin{aligned}
		\gamma_{\parallel} = C_{2} + C_{3}S_{0}^{2} - C_{1}S_{0}
	\end{aligned}
\end{equation}
can become small. In this case a splay deformation can generate a long-lived polar response parallel to the background nematic field.

\section{Director alignment to gradients in scalar order.}
\subsection{$\mathfrak{p} = \mathfrak{q}$}

At the defect core, the orientation of the director field is undefined. To accommodate this in a continuous field, the scalar order parameter at the core of a defect drops to zero. Thus defect cores are generally associated with a high gradient in the scalar order parameter. Defect cores are generally associated with strong spatial variations in the scalar order parameter. These variations can couple to the director orientation through non-reciprocal terms, producing local alignment preferences.

%This drives an alignment of the director around the defect core, which leads to the dominant defect shapes we observe in our simulated results. 

As shown above, the director field is governed by the hydrodynamic equation
\begin{equation}
	\begin{aligned}\label{eq:DA_p=q}
		\partial_{t} f_{\mathfrak{p}} = & A(B-|f_\mathfrak{p}|^2)f_{\mathfrak{p}} + K \Delta  f_{\mathfrak{p}} + \lambda_1 \bigg[ f_{\mathfrak{p}} \cdot \partial^{\mathfrak{p}}_{-} f_{\mathfrak{p}}   - f_{\mathfrak{p}} \cdot \partial^{\mathfrak{p}}_{+} f_{-\mathfrak{p}} \bigg] +  \lambda_2 f_{- \mathfrak{p}}\cdot\partial_+^\mathfrak{p}f_\mathfrak{p} 
	\end{aligned}
\end{equation}

To examine the local alignment mechanism, we consider a one-dimensional variation in the scalar order parameter along the $x$-axis and assume that the director orientation is spatially homogeneous.
%Without loss of generality, we assume that the gradients in the scalar order parameter are parallel to the $x$ axis and write
\begin{equation}
	f_p = S(x)\exp(\mathrm{i}\mathfrak{p}\theta).
\end{equation}
We will consider orientationally ordered systems in which $\theta$ is homogeneous. 

In order to separate the dynamics into parts which affect the orientation and scalar order parameter, we multiply Eq.~\ref{eq:DA_p=q} by $\exp(-\mathrm{i}\mathfrak{p}\theta)$. This gives 
\begin{equation}
	\exp(-\mathrm{i}\mathfrak{p}\theta)\partial_{t}f_p = \partial_{t}S + \mathrm{i}S \mathfrak{p}\partial_{t}\theta.
\end{equation}
Thus the imaginary parts of the resulting dynamical equation describe rotation of the director while the real parts describe changes in the scalar order.

Combining the above we can write the dynamics for the global orientation as
\begin{equation}
	\begin{aligned}\label{eq:DA_theta_p=q}
		\partial_{t} \theta = & \frac{2\lambda_{1} - \lambda_{2}}{\mathfrak{p}} \sin(\mathfrak{p}\theta) \cdot \partial_{x}^\mathfrak{p}S.
	\end{aligned}
\end{equation}

The above equation has stationary solutions at $\theta = n\pi/\mathfrak{p}$, $n \in \mathbb{Z}$

\subsubsection{$\mathfrak{p}=\mathfrak{q}=1$}

We consider a scalar order parameter increasing with $x$, hence $\partial_xS>0$. The stationary states are given by $\theta = 0,\pi$. The stability is given simply by
\begin{equation}
	\begin{aligned}
		2\lambda_{1}-\lambda_{2}>0 & \quad \Rightarrow \quad \theta = \pi \textrm{ is stable}\\
		2\lambda_{1}-\lambda_{2}<0 & \quad \Rightarrow \quad \theta = 0 \textrm{ is stable}\\
	\end{aligned}
\end{equation}

Thus for $2\lambda_{1}-\lambda_{2}>0$ as in our simulations, we expect the director to point down the gradient in $S$. This would be to point toward a defect core, which is exactly as we observe in our simulations. 

\subsubsection{$\mathfrak{p}=\mathfrak{q}=2$}

In this situation, the director orientation couples to the second gradients in $S$. We consider a region in which $\partial_x^2S<0$. The stationary states are given by $\theta = 0,\frac{\pi}{2}$ with stability
\begin{equation}
	\begin{aligned}
		2\lambda_{1}-\lambda_{2}>0 & \quad \Rightarrow \quad \theta = 0 \textrm{ is stable}\\
		2\lambda_{1}-\lambda_{2}<0 & \quad \Rightarrow \quad \theta = \frac{\pi}{2} \textrm{ is stable}\\
	\end{aligned}
\end{equation}

Thus, for negative curvature, the local coupling favors alignment parallel to amplitude variations when \(2\lambda_1-\lambda_2>0\), and perpendicular alignment for the opposite sign. This is motivated by the radial curvature of a passive defect profile, which is negative near the core and asymptotically far from it. However, the calculation assumes a spatially homogeneous director and therefore does not determine the configuration of a full two-dimensional defect. The complete response is calculated in subsequent sections. 
%Thus, for $2\lambda_1-\lambda_2>0$ and $\partial_x^2S>0$, the director preferentially aligns perpendicular to the direction of amplitude variation. However, around a passive topological defect, while the scalar order parameter increases monotonically away from the core, its second radial derivative need not be positive. Consequently, this local alignment argument does not by itself determine the preferred defect configuration. The response of the director starting from a passive defect profile is considered explicitly in subsequent calculations.

%Thus for $2\lambda_{2}-\lambda_{1}>0$ as in our simulations, we expect the director to arrange itself perpendicular to second order variations/second order gradients in $S$. Around a topological defect, the gradients in $S$ are predominantly radial. This would amount to arranging azimuthally around a defect, in a way that promotes bend, which is exactly what we observe in our simulations. 

%Interestingly, it is also the opposite to what would expect given the linear stability analysis shown in the previous section. In other words, this local defect-core alignment need not coincide with the bend-splay preference of the homogeneous ordered state obtained from the linear stability analysis, since the two effects are controlled by different combinations of the continuum coefficients.

%Interestingly, this local alignment preference can differ from the bend–splay relaxation preference obtained from the linear stability analysis in the previous section.

\subsection{$\mathfrak{p}=2\mathfrak{q}=2$}

This system is governed by Eqs.~\ref{eq:f2_21}-\ref{eq:f1_21}.

We consider the case where the primary mode (spin mode) is fully ordered, but the secondary mode (vision cone mode) is fully disordered. Hence 
\begin{equation}
	\begin{aligned}
		f_2 &= S(x)\exp(\mathrm{i}2\theta)\\
		f_1 &= 0
	\end{aligned}
\end{equation}

We consider only the dynamics of $f_2$ given by Eq.~\ref{eq:f2_21} with $f_1=0$ (we revisit this assumption later), which reduce to
\begin{equation}
	\begin{aligned}
		\partial_{t} f_{2} = & A(B-|f_{ 2}|^2)f_{2} + K   \Delta  f_{2} - \lambda_2\partial_{+} f_{2}\cdot \partial_{+} f_{-2}.
	\end{aligned}
\end{equation}

As above, we multiply by $\exp(-\mathrm{i}2\theta)$ and isolate the imaginary part to obtain an equation for the dynamics of the orientation given by
\begin{equation}
	\begin{aligned}
		\partial_{t}\theta = \frac{\lambda_{2}(\partial_{x}S)^2}{2S}\sin(2\theta).
	\end{aligned}
\end{equation}

In this situation, the director orientation couples to the squared gradients in $S$. The stationary states are given by $\theta = 0,\pi/2$ with stability
\begin{equation}
	\begin{aligned}
		\lambda_{2}>0 & \quad \Rightarrow \quad \theta = \frac{\pi}{2} \textrm{ is stable}\\
		\lambda_{2}<0 & \quad \Rightarrow \quad \theta = 0 \textrm{ is stable}.\\
	\end{aligned}
\end{equation}

%Thus, for $\lambda_2>0$, the direct coupling favors alignment perpendicular to gradients in $S$, whereas for $\lambda_{3}$ it favors parallel alignment. The near-onset microscopic derivation gives $\lambda_3>0$ with our coefficient convention. Therefore, this direct contribution alone does not explain the splay-dominated defect configurations observed in our simulations. The complete defect dynamics also involve the secondary polar field $f_{1}$, whose generation and feedback are examined in the following section.

Thus, for $\lambda_2>0$, the direct coupling favors alignment perpendicular to gradients in $S$, whereas for $\lambda_2<0$, it favors parallel alignment. Since this coupling depends on $(\partial_xS)^2$, the alignment preference is independent of the sign of the scalar-order gradient. The near-onset microscopic derivation gives $\lambda_2>0$, so the direct contribution favors perpendicular alignment. However, this local preference alone does not determine the final defect configuration. In particular, the splay-dominated defects observed in the $\{2,1\}$ simulations also involve the generation and feedback of the secondary polar field $f_1$, as examined in subsequent sections.

\section{Defect shape and behavior}
\subsection{Behavior of +1 defects ($\mathfrak{p}=1$)}
\subsubsection{Set Up of the System : Coordinates and Texture}

%We consider the dynamics of the field on a circular shell of fixed radius $r = r_{0}$. On this shell, we neglect the radial variations of the fields. We neglect radial derivatives of the field when evaluating the dynamics on the shell. Using the polar-coordinate definitions, 
In the following defect calculations, we consider both constant and spatially varying scalar-order profiles, depending on the approximation employed. For the $+1$ defects, we consider a circular shell away from the defect core and approximate the scalar order parameter as constant, $S=S_0=\sqrt{B}$, neglecting radial amplitude variations. For the $+1/2$ defects, we instead retain the stationary passive radial profile $S(r)$, including its spatial derivatives. This profile provides the reference configuration from which we evaluate the initial non-reciprocal response. The resulting dynamics can modify both the orientation and the scalar order parameter, so $S(r)$ is not generally assumed to remain fixed in time. Using the polar-coordinate definitions, 
\begin{equation}
	\begin{aligned}
		& \partial_{\pm} = e^{\pm \mathrm{i} \phi} \left( \partial_{r} \pm \frac{\mathrm{i}}{r} \partial_{\phi}\right) = \pm \frac{\mathrm{i} e^{\pm \mathrm{i} \phi}}{r_{0}} \partial_{\phi} \\
		& \Delta = \partial^{2}_{r} + \frac{1}{r} \partial_{r} + \frac{1}{r^{2}}\partial^{2}_{\phi} = \frac{1}{r^{2}_{0}} \partial^{2}_{\phi} \\
		& \partial^{2}_{\pm} = e^{\pm 2 \mathrm{i}\phi} \left[ \partial^{2}_{r} - \frac{1}{r} \partial_{r} - \frac{1}{r^{2}} \partial^{2}_{\phi} \pm \frac{2\mathrm{i}}{r} \partial_{r} \partial_{\phi} \mp \frac{2\mathrm{i}}{r^{2}} \partial_{\phi} \right] = -\frac{e^{\pm 2\mathrm{i} \phi}}{r^{2}_{0}} \left( \partial^{2}_{\phi} \pm 2\mathrm{i} \partial_{\phi} \right)
	\end{aligned}
\end{equation}
where the subsequent equality represents expressions used in an on-shell calculation where $S$ is assumed to be fixed.

We introduce the orientation relative to the local radial direction, 
\begin{equation}
	\begin{aligned}
		\alpha_{n}(\phi,t) = \theta_{n}(\phi,t) - \phi
	\end{aligned}
\end{equation}
such that $\theta_{n} = \phi + \alpha_{n}$. Hence, 
\begin{equation}
	\begin{aligned}
		&\partial_{t} \theta_{n} = \partial_{t} \alpha_{n} \\
		&\partial_{\phi}\theta_{n} = 1+\partial_{\phi} \alpha_{n} \\
		&\partial^{2}_{\phi} \theta_{n} = \partial^{2}_{\phi} \alpha_{n} \\
	\end{aligned}
\end{equation}
We consider systems where $\alpha_n$ is a constant. One thus has $ \partial_{\phi} \theta_{n} = 1 $, which corresponds to a rotationally symmetric $+1$ defect with phase  $\alpha_n$.
% is homogeneous. This restricts the system to one with a $+1$ defect at the origin that is rotationally symmetric, where $\alpha_{n}$ is a constant phase. In the laboratory frame this corresponds to 
% \begin{equation}
	%     \begin{aligned}
		%         \theta_{n}(\phi) = \phi + \alpha_{n}
		%     \end{aligned}
	% \end{equation}
% Therefore, 
% \begin{equation}
	%     \begin{aligned}
		%         \partial_{\phi} \theta_{n} = 1 
		%     \end{aligned}
	% \end{equation}
% so that the uniform state in the circular coordinate system naturally corresponds to a $+1$ winding in the laboratory frame. 
% The corresponding Fourier mode is
% \begin{equation}
	%     \begin{aligned}
		%         f_{n} = A_{n} e^{\mathrm{i}n(\phi + \alpha_{n})}
		%     \end{aligned}
	% \end{equation}
% Thus, the constant phase, $\alpha_{n}$ determines the phase of the defect. 
In particular, $\alpha_{n}=0$ corresponds to an aster; while $\alpha_{n}=\frac{\pi}{2}$ corresponds to a vortex. Intermediate values correspond to spiral configurations.
Far from the defect core, the amplitude of mode $n$ is $A_n=\sqrt{B}$.

Using the above change in coordinate system and variables, the dynamical equations for the $\mathfrak{p}=1$ systems can be re-written to describe the behavior of the singular defect phase $\alpha_n$.

\subsubsection{$+1$ Defect phase for $\mathfrak{p}=1;\mathfrak{q}=1$}

At large radii ($r_0$), the dynamical equation for the $\mathfrak{p}=1$, $\mathfrak{q}=1$ system can be reduced to:
\begin{equation}
	\begin{aligned}
		\partial_{t} \alpha  = \frac{(\lambda_{2}+2\lambda_{1})S_{0}}{r_{0}} \sin (\alpha)
	\end{aligned}
\end{equation}

%\ylann{you could recall what is $S_0$ and how you obtained this equation. }
When $(\lambda_{2} +2\lambda_{1}) \neq 0$ the above equation has stationary states when $\alpha=n\pi$; $n\in\mathbb{Z}$. The stability of these states depends on the sign of $(\lambda_{2}+ 2\lambda_{1})$ giving 
% \ylann{By taking the definitions of $\lambda_{1,2}$  , one has $ 3\lambda_{2} -\lambda_{1}<0 \Leftrightarrow D<D_c/6$ with the critical temperature $D_c  = Jf_0/(2\p)$, in the case $\sigma>0$. How do we rationalise this qualitative change of behaviour as the temperature is varied ? I've never seen this in my simulations. Do you ? }
\begin{equation}
	\begin{aligned}
		& (\lambda_{2} +2\lambda_{1}) < 0 \quad \Rightarrow \quad \alpha = 2n\pi \ : \ \textrm{sources (= outwards asters) are the only stable configurations} \\
		& (\lambda_{2} +2\lambda_{1}) > 0 \quad \Rightarrow \quad \alpha = (2n+1)\pi \ : \ \textrm{sinks (= inwards asters) are the only stable configurations} \\
	\end{aligned}
\end{equation}
This is consistent with previous work published on the $\{1,1\}$ model \cite{vafa2022defect, rouzaire2025nonreciprocal} and the simulations presented in the main text. 

\subsubsection{$+1$ Defect phase for $\mathfrak{p}=1;\mathfrak{q}=2$}
We following a similar approach (and assume large $r_{0}$)  to write a dynamical equation for the defect phase of a $+1$ defect in the $\{1,2\}$ system as
\begin{equation}
	\begin{aligned}
		\partial_{t} \alpha = - \frac{\lambda_{1}}{r^{2}_{0}} \sin(2\alpha) 
	\end{aligned}
\end{equation}
The defect has two classes of stationary textures when $\lambda_1\neq0$, which are 
\begin{equation}
	\begin{aligned}
		& \lambda_{1} > 0 \quad \Rightarrow \quad \alpha = n\pi \ : \ \textrm{sources and sinks are the only stable configurations} \\
		& \lambda_{1}<0 \quad \Rightarrow \quad \alpha = \frac{(2n+1)\pi}{2} \ : \ \textrm{vortices are the only stable configuration} \\
	\end{aligned}
\end{equation}

Interestingly, the $\{1,2\}$ system distinguished between bend and splay, reflecting the nematic nature of the vision cones. Conversely, the $\{1,1\}$ system distinguishes between convergence and divergence reflecting its polar only nature. 

\subsubsection{Gradient Dynamics of the $\{1,2\}$ system.}

The above result can be explained by more carefully analyzing the coarse-grained dynamics of the $\{1,2\}$ system given by
\begin{equation} \label{EQNR190}
	\begin{aligned}
		\partial_{t} f_{1} = A(B-|f_{1}|^{2})f_{1} + K \Delta f_{1} - \lambda_{1} \partial^{2}_{+} f_{-1}
	\end{aligned}
\end{equation}
and we can show easily that the non-reciprocal term in Eq.~\ref{EQNR190} has potential dynamics. In particular, 
\begin{equation}
	\begin{aligned}
		\partial_{t} f_{1} = - \frac{\delta \mathcal{F}}{\delta f_{-1}} 
	\end{aligned}
\end{equation}
where 
\begin{equation} \label{EQNR051}
	\begin{aligned}
		\mathcal{F} = \int d^{2} \mathbf{r} \left[- AB |f_{1}|^{2} + \frac{A}{2} |f_{1}|^{4} + K (\nabla f_{1}) (\nabla f_{-1}) + \frac{\lambda_{1}}{2} \left( f_{-1} \partial^{2}_{+} f_{-1} + f_{1} \partial^{2}_{-} f_{1}\right)\right] 
	\end{aligned}
\end{equation}
Note: the dynamics of $f_{1}$ stem from the derivative of $\mathcal{F}$ with respect to $f_{-1}$, just as the dynamics of a covariant vector derive from the derivative with respect to a contravariant vector in curved spaces. Another way to convince oneself is note that $\mathcal{F}$ is a real number and $f_{\pm1}$ are complex conjugate of one another. The only way to satisfy all these conditions is precisely to have the structure $\partial_{t} f_{p} = -\delta \mathcal{F}/\delta f_{-p} $. 
\newline 

Since one has a gradient dynamics,
\begin{equation}
	\begin{aligned}
		\frac{d\mathcal{F}}{dt}\leq0 
	\end{aligned}
\end{equation}
and the continuum dynamics correspond to gradient descent of the effective free energy. To make the physical meaning of the $\lambda_{1}-$term explicit, we write the polar mode as
\begin{equation}
	\begin{aligned}
		f_{1}=P_x+\mathrm{i}P_y  \quad ; \quad \mathbf{P}=(P_x,P_y)
	\end{aligned}
\end{equation}
Then
\begin{equation}
	\begin{aligned}
		\partial_{-}f_{1} &= \nabla\cdot\mathbf{P} + \mathrm{i} \left(\nabla\times\mathbf{P}\right) \\
		\partial_{+}f_{-1} &= \nabla\cdot\mathbf{P} - \mathrm{i}\left(\nabla\times\mathbf{P}\right).
	\end{aligned}
\end{equation}
Using integration by parts, Eq.~\ref{EQNR051} can therefore be written as
\begin{equation}\label{EQNR052}
	\begin{aligned}
		\mathcal{F} = \int d^{2}\mathbf{r} \bigg[ &-AB|\mathbf{P}|^{2} + \frac{A}{2}|\mathbf{P}|^{4} \\ 
		&+ \left(K-\lambda_{1}\right) \left(\nabla\cdot\mathbf{P}\right)^{2} + \left(K+\lambda_{1}\right) \left(\nabla\times\mathbf{P}\right)^{2} \bigg]
	\end{aligned}
\end{equation}

Thus the $\lambda_{1}-$term does not generate a genuinely non-potential contribution at this order, but instead separates the elastic cost of splay and bend. The corresponding gradient coefficients are
\begin{equation}\label{EQNR053}
	\begin{aligned}
		K_{\textrm{splay}} &= K-\lambda_{1} \\
		K_{\textrm{bend}} &= K+\lambda_{1}
	\end{aligned}
\end{equation}
Equivalently, the polar-vector dynamics may be written as
\begin{equation}\label{EQNR054}
	\begin{aligned}
		\partial_t\mathbf{P} = A\left(B-|\mathbf{P}|^{2}\right)\mathbf{P} + \left(K-\lambda_{1}\right) \nabla\left(\nabla\cdot\mathbf{P}\right) - \left(K+\lambda_{1}\right) \nabla\times \left[ \left(\nabla\times\mathbf{P}\right)
		\right]
	\end{aligned}
\end{equation}
%\ylann{This is weird... The grad div term can be rewritten with a curl curl \\
	%see \href{https://en.wikipedia.org/wiki/Vector_calculus_identities#Second_derivatives}{(link)} \\
	%such that the last 2 terms add up to  
	%\begin{equation}
	%    -2\lambda_1 \nabla\times \left(\nabla\times\mathbf{P}
	%        \right)  + (K-\lambda_{1}) \Delta \textbf{P}
	% \end{equation}
% While the laplacian $\Delta \textbf{P}$ is an equilibrium term, the other is not (I had the same in refs \cite{rouzaire2025nonreciprocal, rouzaire2026dynamics} for the (1,1) case). Yet, we say that the whole dynamics comes from a gradient. Where is my/the error ? 
% } 

The homogeneous ordered state is stable to both distortions provided
\begin{equation}
\begin{aligned}
	K>|\lambda_{1}|
\end{aligned}
\end{equation}
For $\lambda_{1}>0$, splay is the lower-energy distortion while bend is more strongly penalized and relaxes faster. Conversely, for $\lambda_{1}<0$, bend is the low-energy distortion and splay is more strongly penalized. Therefore, at the order considered, the effect of the non-reciprocal coupling in the $\mathfrak{p}=1;\mathfrak{q}=2$ Model is completely captured by a conservative splitting of the bend and splay elastic coefficients.

\subsection{$+\frac{1}{2}-$Defect in the $\mathfrak{p}=2;\mathfrak{q}=2$ Model}

The hydrodynamic equation describing the $\mathfrak{p}=2;\mathfrak{q}=2$ system is given by 
\begin{equation}\label{2201}
\begin{aligned}
	\partial_{t} f_{2} = A\left(B-|f_{2}|^{2}\right)f_{2} + K \Delta f_{2} + \lambda_{1} f_{2} \left[ \partial^{2}_{-} f_{2} - \partial^{2}_{+} f_{-2} \right] + \lambda_{2} f_{-2} \partial^{2}_{+} f_{2} 
\end{aligned}
\end{equation}

Writing
\begin{equation}
\begin{aligned}
	f_{2}=Se^{2\mathrm{i}\theta}
\end{aligned}
\end{equation}
we get
\begin{equation}
\begin{aligned}
	e^{-2\mathrm{i}\theta}\partial_{t} f_{2} = \partial_{t}S + 2\mathrm{i}S\partial_{t} \theta.
\end{aligned}
\end{equation}
Thus, the real part of $e^{-2\mathrm{i}\theta}\partial_{t} f_{2}$ controls the scalar order $S$, while the imaginary part controls the orientation $\theta$. We use this to evaluate the action of the non-reciprocal parts of Eq.~\ref{2201}.  The $\lambda_{1}$ term gives
\begin{equation}
\begin{aligned}
	e^{-2\mathrm{i}\theta} \cdot \lambda_{1} f_{2} \left[
	\partial^{2}_{-}f_{2} -\partial^{2}_{+}f_{-2} \right] \equiv S\cdot(\textrm{Imaginary Number})
\end{aligned}
\end{equation}
since
\begin{equation}
\begin{aligned}
	\partial^{2}_{+}f_{-2}=
	\left(\partial^{2}_{-}f_{2}\right)^{*}
\end{aligned}
\end{equation}
%Hence,
%\begin{equation}
%    \begin{aligned}
%        \partial_{t}S\bigg|_{\beta,\textrm{direct}}=0
%    \end{aligned}
%\end{equation}
The $\lambda_{1}-$operator is thus exactly orientation-only in its direct action: it moves the field along the constant-$S$ manifold.

For the $\lambda_{2}-$operator,
\begin{equation}
\begin{aligned}
	e^{-2\mathrm{i}\theta}\cdot\lambda_{2} f_{-2}\partial^{2}_{+}f_{2}=\lambda_{2} S e^{-4\mathrm{i}\theta}\partial^{2}_{+}f_{2},
\end{aligned}
\end{equation}
which is generically complex and thus affects both $S$ and $\theta$.

\subsubsection{Defect Self-Propulsion}
We now consider the dynamics of the field around a $+\frac{1}{2}$ defect. We assume the initial shape of the defect to be that which satisfies the passive parts of Eq.~\ref{2201}, that is $\theta=\phi/2 + \theta_0$. $\theta_0$ is the phase associated with the defect, which controls its orientation; without loss of generality we set $\theta_0=0$, which leaves us with the field
\begin{equation}
\begin{aligned}
	f^{+\frac{1}{2}}_{2}(r,\phi)=S(r)e^{\mathrm{i}\phi}.
\end{aligned}
\end{equation}
Since the configuration $f^{+\frac{1}{2}}_{2}$ is a solution to the passive parts of Eq.~\ref{2201}, the initial dynamics of the defect are dictated by the non-reciprocal parts governed by $\lambda_{1}$ and $\lambda_{2}$. In radial coordinates, these terms are given by 
%where
%\begin{equation}
%    \begin{aligned}
%        K\left(\partial^{2}_{r}S+\frac{\partial_{r}S}{r} - \frac{S}{r^{2}}\right)+ A(B-S^{2})S=0 \equiv \textrm{Radial Amplitude Passive Relation}
%    \end{aligned}
%\end{equation}
%Since the passive terms vanish on $f^{+\frac{1}{2}}_{2}$, the non-linear terms generate the residual forcing
\begin{equation}\label{2202}
\begin{aligned}
	R = \lambda_{1} S\mathcal{B}(r)\left[1-e^{2\mathrm{i}\phi}\right] + \lambda_{2} S\mathcal{A}(r)e^{2\mathrm{i}\phi}
\end{aligned}
\end{equation}
where
\begin{equation}
\begin{aligned}
	\mathcal{A}(r) &= \partial^{2}_{r}S-\frac{3\partial_{r}S}{r} + \frac{3S}{r^{2}} \\
	\mathcal{B}(r) &= \partial^{2}_{r}S + \frac{\partial_{r}S}{r} - \frac{S}{r^{2}}
\end{aligned}
\end{equation}
As above, we can separate the dynamics into changes in $\theta$ and $S$ by multiplying $R$ by $e^{-2\mathrm{i}\theta}=e^{-\mathrm{i}\phi}$ which gives 
%Since the passive defect has phase $f^{+\frac{1}{2}}_{2}=Se^{\mathrm{i}\phi}$, it is more useful to express the forcing relative to the local defect phase. Multiplying by $e^{-\mathrm{i}\phi}$ gives
\begin{equation}\label{EQNR062}
\begin{aligned}
	e^{-\mathrm{i}\phi}R &= \lambda_{2} S\mathcal{A}(r)\cos(\phi) + \mathrm{i}S \left[ \lambda_{2}\mathcal{A}(r) - 2\lambda_{1}\mathcal{B}(r) \right] \sin(\phi) \Rightarrow
	\begin{cases}
		\partial_{t}S =
		\lambda_{2} S\mathcal{A}(r)\cos(\phi)\\
		\partial_{t}\theta = \left[ \frac{\lambda_{2}}{2}\mathcal{A}(r) - \lambda_{1}\mathcal{B}(r)\right]\sin(\phi)
	\end{cases}
\end{aligned}
\end{equation}
Equation \ref{EQNR062} gives the instantaneous response of a neutrally shaped $+\frac{1}{2}-$defect to the non-reciprocal terms. Some primary observations: It shows that the forcing around a $+\frac{1}{2}-$defect is intrinsically polar, reflecting the shape of the defect. 

The $\lambda_{2}-$operator leads to a front-back asymmetry in the scalar order parameter. The orientation response also reflects the polar symmetry of the $+\frac{1}{2}-$defect, leading to a front-back asymmetry in the distribution of orientational gradients. 

To highlight the effect of the orientational dynamics around the defect, we decompose the dynamics into a rigid translation of the defect and a shape deformation. Written explicitly, for small times $t$ we expect 
\begin{equation}
\begin{aligned}
	f_{2}(\mathbf{r},t) \simeq f^{+\frac{1}{2}}_{2} \left( \mathbf{r} - t{\mathbf{v}} \right) + \delta f_{\textrm{shape}}
\end{aligned}
\end{equation}
where ${\mathbf{v}}$ is the initial velocity of self propulsion. 

When we set the phase of the $+\frac{1}{2}-$defect to zero ($\theta_0=0$) we restricted its axis of symmetry to align with the $x-$axis, which fixes the possible direction of translation. Letting $\mathbf{v} = v_{1/2}\hat{\mathbf{x}}$ and taking the time derivative of the above equation we arrive at
\begin{equation}
\begin{aligned}
	\partial_{t}f_{2} = -v_{1/2}\partial_xf^{+\frac{1}{2}}_{2} + \partial_{t} \delta f_{\textrm{shape}}.
\end{aligned}
\end{equation}

%the orientational response in \ref{EQNR062} contains both deformation and rigid translation of the defect and therefore should not by itself be interpreted as a bend-versus-splay shape selection criterion. We first determine the direction of translation through projection onto the translational mode. The internal deformation of the defect is separated from this translation below.

%translational mode of the passive defect for displacement along the polarity axis, which we choose to be the $x-$axis, is
At time $t=0$ we set $f_2 = f^{+\frac{1}{2}}_{2}$, thus the initial dynamics are $\partial_tf_2 = R$ defined in Eq.~\ref{2202} above. At time $t=0$ we also directly calculate
\begin{equation}
\begin{aligned}
	T_x=\partial_{x} f^{+\frac{1}{2}}_{2} = \frac{1}{2} \left[ \left( \partial_{r}S+\frac{S}{r} \right) + \left( \partial_{r}S-\frac{S}{r} \right)e^{2\mathrm{i}\phi} \right]
\end{aligned}
\end{equation}
%This is the crucial observation: the translational mode occupies the same $m=0,2$ harmonic sector as the residual,
%\begin{equation}
%    \begin{aligned}
%        R\in\{m=0,2\} \quad \textrm{and} \quad T_{x}\in\{m=0,2\}
%    \end{aligned}
%\end{equation}
%The non-linear forcing therefore has a finite projection onto the translation mode of the $+\frac{1}{2}-$defect. For comparison, translation perpendicular to the defect polarity gives
%\begin{equation}
%    \begin{aligned}
%        T_{y} \equiv
%        \partial_{y}f^{+\frac{1}{2}}_{2} = \frac{\mathrm{i}}{2} \left[ \left( \partial_{r}S+\frac{S}{r} \right) - \left( \partial_{r}S-\frac{S}{r} \right)e^{2\mathrm{i}\phi} \right]
%    \end{aligned}
%\end{equation}
%For the chosen defect phase, $2\theta=\phi$, symmetry eliminates the projection of the residual $R$ onto this transverse mode. Hence the non-linear terms select motion along the polar axis of the defect.

%To make the resulting velocity explicit, we write the slowly moving defect as
%\begin{equation}
%    \begin{aligned}
%        f_{2}(\mathbf{r},t) \simeq f^{+\frac{1}{2}}_{2} \left( \mathbf{r} - X(t)\hat{\mathbf{x}} \right) + \delta f_{\textrm{shape}}
%    \end{aligned}
%\end{equation}
%Then,
%\begin{equation}
%    \begin{aligned}
%        \partial_{t}f_{2} = -\dot{X}T_x + \partial_{t} \delta %f_{\textrm{shape}}
%    \end{aligned}
%\end{equation}
We estimate the defect velocity by projecting the dynamics onto the translational zero mode, which removes the defect shape deformations. 

We define the real inner product $\langle u,v\rangle \equiv \textrm{Re} \left[\int d^{2} \mathbf{r} \  u^{*}v\right]$ which allows us to write,
\begin{equation}\label{EQNR063}
\begin{aligned}
	v_{1/2} &= -
	\frac{\pi \int_{0}^{\infty} rdr \ \left[\lambda_{2} S \left( \partial_{r}S-\frac{S}{r} \right) \mathcal{A}(r) + \frac{2\lambda_{1} S^{2}}{r} \mathcal{B}(r) \right]}{\frac{\pi}{2}\int_{0}^{\infty} rdr \ \left[ \left( \partial_rS+\frac{S}{r} \right)^{2} + \left( \partial_rS-\frac{S}{r} \right)^{2} \right] }%\\ 
	%&= - \frac{\langle T_{x},R\rangle}{\langle T_{x},T_{x}\rangle}
\end{aligned}
\end{equation}
For an isolated two-dimensional defect, the translational norm is logarithmically dependent on the outer system size. The radial integrals in Eq.~\ref{EQNR063} and below are therefore understood with the finite outer cutoff of the physical or numerical system.
\newline
\newline
It is useful to separate the two contributions to this velocity and write
\begin{equation}\label{EQNR064}
\begin{aligned}
	v_{\frac{1}{2}} = C_{\lambda_{1}}\lambda_{1} + C_{\lambda_{2}}\lambda_{2}
\end{aligned}
\end{equation}
where
\begin{equation}\label{EQNR065}
\begin{aligned}
	C_{\lambda_{1}} & = - \frac{2\pi}{\langle T_x,T_x\rangle } \int_{0}^{\infty} dr \ S^{2}\mathcal{B}(r) \\
	C_{\lambda_{2}} & = - \frac{\pi}{ \langle T_{x},T_{x}\rangle } \int_{0}^{\infty}
	rdr \  S \left( \partial_{r}S-\frac{S}{r} \right) \mathcal{A}(r)
\end{aligned}
\end{equation}
For our initial conditions we can write
\begin{equation}
\begin{aligned}
	S(r)\geq0; \quad \partial_{r}S\geq0; \quad S(r)\leq\sqrt{B}
\end{aligned}
\end{equation}
For the $\lambda_{2}-$contribution,
\begin{equation}
\begin{aligned}
	\int_{0}^{\infty} r \ dr \ S \left( \partial_{r}S-\frac{S}{r} \right) \mathcal{A}(r) = - \frac{1}{2} \int_{0}^{\infty} dr \ \left( r\partial_{r}S+5S \right) \left( \partial_{r}S-\frac{S}{r} \right)^{2} <0
\end{aligned}
\end{equation}
Hence,
\begin{equation}
\begin{aligned}
	C_{\lambda_{2}} = \frac{\pi}{2\langle T_{x},T_{x}\rangle} \int_{0}^{\infty} dr \  \left( r\partial_{r}S+5S \right) \left( \partial_{r}S-\frac{S}{r} \right)^{2} >0 
\end{aligned}
\end{equation}
Similarly, the passive radial relation gives
\begin{equation}
\begin{aligned}
	\mathcal{B}(r) = -\frac{A}{K} \left( B-S^{2} \right)S
\end{aligned}
\end{equation}
so that
\begin{equation}
\begin{aligned}
	C_{\lambda_{1}} = \frac{ 2\pi A }{ K\langle T_x,T_x\rangle } \int_{0}^{\infty} dr\  \left( B-S^{2} \right)S^{3} >0
\end{aligned}
\end{equation}
Thus, for the phase convention used here, where the positive $x-$direction is taken to be the tail direction,
\begin{equation}
\begin{aligned}
	C_{\lambda_{1}}\lambda_{1} + C_{\lambda_{2}}\lambda_{2}>0 & \quad \Rightarrow \quad \textrm{motion towards the tail} \\
	C_{\lambda_{1}}\lambda_{1} + C_{\lambda_{2}}\lambda_{2}<0 & \quad \Rightarrow \quad \textrm{motion towards the head}
\end{aligned}
\end{equation}
From our microscopic model, we expect $\lambda_{1} > 0$, which would drive motion towards the tail as observed. However, it is possible for $\lambda_{2}<0$, i.e., $\lambda_{2}$ to be negative, which provides an avenue to defects propelling in the opposite direction. While we do not see this in the microscopic model, it is confirmed in our hydrodynamic simulations. 

\subsubsection{Changing defect shape}

In both our lattice and hydrodynamic simulations of the $\{2,2\}$ system, we observe that the head of a $+1/2$ defect becomes bend dominated, with the splay being almost entirely removed. This actually contradicts what is suggested by our linear stability analysis, but rather emerges dynamically in the vicinity of a defect. The following calculation allows us to understand the changing defect shape. 

The residual $R$ calculated in Eq.~\ref{2202} is the full instantaneous response of the initial $+1/2$ defect to the non-linear continuum terms. Importantly, this residual contains both the part which translates the defect and the part which changes its internal structure. We therefore write
\begin{equation}\label{EQNR068}
\begin{aligned}
	R = R_{\textrm{translation}} + R_{\textrm{shape}}.
\end{aligned}
\end{equation}
For a defect translating with velocity $v_{+\frac{1}{2}}$ along the $x-$axis,
\begin{equation}
\begin{aligned}
	R_{\textrm{translation}} = -v_{+\frac{1}{2}}\partial_{x}f_2^{+\frac{1}{2}}
\end{aligned}
\end{equation}
Consequently,
\begin{equation}\label{EQNR069}
\begin{aligned}
	R_{\textrm{shape}} = R + v_{+\frac{1}{2}}\partial_{x}f_2^{+\frac{1}{2}}
\end{aligned}
\end{equation}
The velocity in Eq.~\ref{EQNR063} is chosen precisely such that
\begin{equation}
\begin{aligned}
	\left\langle \partial_{x}f_2^{+\frac{1}{2}},R_{\textrm{shape}}\right\rangle=0
\end{aligned}
\end{equation}
Thus, $R_{\textrm{shape}}$ contains the part of the leading-order response which cannot be absorbed into a rigid translation of the passive defect. As before, we multiply by $e^{-\mathrm{i}\phi}$ separate changes in amplitude and phase to the real and imaginary parts, respectively. This allows us to write
\begin{equation}
\begin{aligned}
	e^{-\mathrm{i}\phi}\partial_{x}f_2^{+\frac{1}{2}} = \partial_rS\cos(\phi) - \mathrm{i} \frac{S}{r} \sin(\phi),
\end{aligned}
\end{equation}
\begin{equation}
\begin{aligned}
	e^{-\mathrm{i}\phi}R = \lambda_{2} S\mathcal{A}(r)\cos(\phi) + \mathrm{i}S \left[\lambda_{2} \mathcal{A}(r)-2\lambda_{1} \mathcal{B}(r)\right] \sin(\phi).
\end{aligned}
\end{equation}
Which we combine to obtain
\begin{equation}\label{EQNR070}
\begin{aligned}
	e^{-\mathrm{i}\phi} R_{\textrm{shape}} &= \left[ \lambda_{2} S\mathcal{A}(r) + v_{+\frac{1}{2}}\partial_{r}S\right] \cos(\phi)+ \mathrm{i}S \left[ \lambda_{2}\mathcal{A}(r) - 2\lambda_{1}\mathcal{B}(r) - \frac{v_{+\frac{1}{2}}}{r} \right] \sin(\phi)
\end{aligned}
\end{equation}
Therefore, the leading-order translation-subtracted internal response is
\begin{equation}\label{EQNR071}
\begin{aligned}
	\partial_{t}S \bigg|_{\textrm{shape}} &= \left[ \lambda_{2} S\mathcal{A}(r) + v_{+\frac{1}{2}}\partial_{r}S \right] \cos(\phi) \\
	\partial_{t}\theta \bigg|_{\textrm{shape}} &= \left[ \frac{\lambda_{2}}{2}\mathcal{A}(r) - \lambda_{1}\mathcal{B}(r) - \frac{v_{+\frac{1}{2}}}{2r} \right] \sin(\phi)
\end{aligned}
\end{equation}

Since we are concerned with the changing distribution of bend and splay, we will focus entirely on the dynamics of the orientation field, given by the second equation above. For convenience we introduce the following definition
\begin{equation}
\delta(r) =  \frac{\lambda_{2}}{2}\mathcal{A}(r) - \lambda_{1}\mathcal{B}(r) - \frac{v_{+\frac{1}{2}}}{2r}.
\end{equation}

We assume that at small times $t$, the orientation field follows 
\begin{equation}\label{EQNR073}
\begin{aligned}
	\theta(r,\phi)= \frac{\phi}{2} + t\delta(r)\sin(\phi)
\end{aligned}
\end{equation}

Using this definition we estimate the bend and splay fields, given by $\left( \nabla\times\mathbf{n} \right)^{2}$ and $\left( \nabla\cdot\mathbf{n} \right)^{2}$, respectively. Around a $+1/2$ defect, the sum of bend and splay is conserved, thus by evaluating $\left( \nabla\cdot\mathbf{n} \right)^{2} - \left( \nabla\times\mathbf{n} \right)^{2}=0$ we can evaluate the angle at which the distortions switch between bend and splay dominated regions. 

From Eq.~\ref{EQNR073}, we evaluate to linear order in $t$
\begin{equation}\label{EQNR074}
\begin{aligned}
	0=\left( \nabla\cdot\mathbf{n} \right)^{2} - \left( \nabla\times\mathbf{n} \right)^{2} = \frac{1}{r^{2}} \left[ \frac{\cos(\phi)}{4} + \frac{t\delta(r)}{2} \left( 1+\cos^{2}\phi \right) + tr\partial_{r}\delta(r) \sin^{2}\phi \right] +
	\mathcal{O}(t^{2}).
\end{aligned}
\end{equation}

At time $t=0$, the above relation is equivalent to $\cos(\phi)=0$, which has solutions at $\phi=\pm\pi/2$. Thus for our initial conditions, the field around the defect has equal bend and splay, as expected. 

For small $t$, we expect the solution to be of the form $\phi = \pi/2 + \delta\phi$. Substituting this into Eq.~\ref{EQNR074} gives to linear order in $\delta\phi$
\begin{equation}
\delta\phi = 4rt\left[\frac{\delta(r)}{2r}+\partial_r\delta(r)\right].
\end{equation}

For $|\lambda_{2}|<|\lambda_{1}|$ and $\lambda_{1}>0$, which captures best the lattice based model, we evaluate $\frac{\delta(r)}{2r}+\partial_r\delta(r)<0$, which indicates that the bend dominated region grows. This is exactly as observed in both lattice and hydrodynamic simulations of the $\{2,2\}$ model. For other values of $\lambda_{1,2}$ the opposite behavior is possible.

\subsection{$+\frac{1}{2}-$Defect in $\mathfrak{p}=2;\mathfrak{q}=1$ Model}

The $\mathfrak{p} = 2;\mathfrak{q}=1$ Model is qualitatively different from the $\mathfrak{p}=2;\mathfrak{q}=2$ Model because the critical nematic mode $f_{2}$ is accompanied by a secondary slow polar mode $f_{1}$. The reduced-continuum equations are 
\begin{equation}\label{2101}
\begin{aligned}
	\partial_{t} f_{2\mathfrak{q}} = & A(B-|f_{ 2\mathfrak{q}}|^2)f_{2\mathfrak{q}} + K   \Delta  f_{2\mathfrak{q}} \\
	& + \lambda_{1} \left[ f_{\mathfrak{q}} \partial^{\mathfrak{q}}_{-} f_{2\mathfrak{q}} + f_{-\mathfrak{q}} \partial^{\mathfrak{q}}_{+} f_{2\mathfrak{q}}\right] - \lambda_{2} \left( \partial^{\mathfrak{q}}_{+} f_{2\mathfrak{q}}\right) \left( \partial^{\mathfrak{q}}_{+} f_{-2\mathfrak{q}} \right) 
\end{aligned}
\end{equation}
\begin{equation}\label{2102}
\begin{aligned}
	\partial_{t} f_{\mathfrak{q}} = & C_{1}f_{2\mathfrak{q}}\cdot f_{-\mathfrak{q}}- C_{2}f_{\mathfrak{q}} - C_{3}|f_{2\mathfrak{q}}|^2f_{\mathfrak{q}}  + \frac{K}{2} f_{-\mathfrak{q}}\Delta f_{2\mathfrak{q}}\\
	& + \frac{\lambda_{1}}{2} \left[ \partial^{\mathfrak{q}}_{-} f_{2\mathfrak{q}} + f_{-2\mathfrak{q}}\partial^{\mathfrak{q}}_{+} f_{2\mathfrak{q}} - f_{2\mathfrak{q}} \partial^{\mathfrak{q}}_{+} f_{-2\mathfrak{q}} \right]
\end{aligned}    
\end{equation}
Here, gradients in $f_{2}$ can generate a secondary polar field Eq.~\ref{2102}, which can subsequently feedback in the nematic field equation $f_{2}$ though Eq.~\ref{2101}. The $\lambda_2$ contribution acts directly on the nematic texture, even when $f_1=0$. Thus, the reduced theory contains both a direct quadratic-gradient mechanism and a polar-mediated mechanism for defect motion and deformation.

Writing $f_{2}=S e^{2\mathrm{i}\theta}$, $f_{1} = P_{x} + \mathrm{i}P_{y}$, the real fields $P_{x}$ and $P_{y}$ are the components of the local polar order. An important feature of Eq.~\ref{2101} is that, when $f_{1}=0$, the $\lambda_{1}$ terms vanish from the nematic equation, $\lambda_{2}$ can act directly on the nematic texture. Thus, in the general $\mathfrak{p}=2;\mathfrak{q}=1$ Model, defect dynamics can contain both a direct nematic contribution and a polar-mediated contribution.

\subsubsection{Defect Induced Polar Order}

As before, we assume an initial condition that satisfies the reciprocal parts of the $f_2$ equation in the presence of a $+\frac{1}{2}-$defect, 
\begin{equation}
\begin{aligned}
	f^{+\frac{1}{2}}_{2} (r,\phi) =  S_{r} e^{\mathrm{i}\phi} \quad ; \quad f_{1} = 0
\end{aligned}
\end{equation}

As before, we separate the dynamical equation into the reciprocal and non-reciprocal parts. The reciprocal parts govern the radial variations in the scalar order parameter around the defect.
\begin{equation}
\begin{aligned}
	K \left( \partial_{r}^{2}S + \frac{\partial_{r}S}{r} - \frac{S}{r^{2}} \right) + A(B-S^{2})S = 0 \equiv \textrm{Radial Amplitude Passive Relation}
\end{aligned}
\end{equation}
As in the $\mathfrak{p}=2;\mathfrak{q}=2$ calculation, we write
\begin{equation}
\begin{aligned}
	\mathcal{B}(r) = \partial_{r}^{2}S + \frac{\partial_{r}S}{r} - \frac{S}{r^{2}}
\end{aligned}
\end{equation}
so that
\begin{equation}
\begin{aligned}
	\Delta f_{2}^{+\frac{1}{2}} = \mathcal{B}(r)e^{\mathrm{i}\phi}
\end{aligned}
\end{equation}
For $\mathfrak{q}=1$,
\begin{equation}\label{EQNR081}
\begin{aligned}
	\partial_{-}f_{2}^{+\frac{1}{2}} = \partial_{r}S+\frac{S}{r} \quad & ; \quad \partial_{-}f_{-2}^{+\frac{1}{2}} = \left( \partial_{r}S-\frac{S}{r} \right)e^{-2\mathrm{i}\phi} \\
	\partial_{+}f_{-2}^{+\frac{1}{2}} = \partial_{r}S+\frac{S}{r} \quad & ; \quad \partial_{+}f_{2}^{+\frac{1}{2}} = \left( \partial_{r}S-\frac{S}{r} \right)e^{2\mathrm{i}\phi} \\
\end{aligned}
\end{equation}
We now evaluate the polar equation Eq.~\ref{eq:f1_21} on the passive defect. Since $f_{1}=0$ initially, all terms proportional to $f_{\pm1}$ vanish and the remaining terms generate an additive polar source. Using Eq.~\ref{EQNR081}, we obtain
\begin{equation}\label{EQNR082}
\begin{aligned}
	\partial_{t}f_{1} \bigg|_{f_{1}=0} &= \left[ \frac{\lambda_{1}f_{0}}{2} \left( \partial_{r}S+\frac{S}{r} \right) \right] -  \lambda_{1}\frac{S^{2}}{r}  e^{\mathrm{i}\phi}
\end{aligned}
\end{equation}
Thus, the passive nematic $+\frac{1}{2}-$defect generates two angular harmonics of the secondary polar field. We therefore write
\begin{equation}\label{EQNR083}
\begin{aligned}
	f_{1}(r,\phi,t) = p_{0}(r,t) + p_{1}(r,t)e^{\mathrm{i}\phi}
\end{aligned}
\end{equation}
For an initially vanishing polar field and real continuum coefficients, $p_{0}$ and $p_{1}$ may be taken to be real at this order. The $p_{0}$ term simply aligns with the orientation of the defect, whereas the $p_{1}$ term generates a $+1$ defect in the $f_1$ field. 
\newline
\newline

The terms in Eq.~\ref{2102} which are linear in $f_{1}$ give
\begin{equation}
\begin{aligned}
	\partial_{t}f_{1} \bigg|_{\textrm{linear in }f_{1}} = - \left[ C_{2} + C_{3}S^{2} \right]f_{1} + \left[ C_{1}S + \frac{K}{2}\mathcal{B}(r) \right] e^{\mathrm{i}\phi}f_{-1}.
\end{aligned}
\end{equation}
Substituting Eq.~\ref{EQNR083} therefore gives
\begin{equation}\label{EQNR084}
\begin{aligned}
	\partial_{t}p_{0} &= - \left[ C_{2} + C_{3}S^{2} \right]p_{0} + \left[ C_{1}S + \frac{K}{2}\mathcal{B}(r) \right]p_{1} + \frac{\lambda_{1}f_{0}}{2} \left( \partial_{r}S+\frac{S}{r} \right) , \\
	\partial_{t}p_{1} &= - \left[ C_{2} + C_{3}S^{2} \right]p_{1} + \left[ C_{1}S + \frac{K}{2}\mathcal{B}(r) \right]p_{0} - \lambda_{1}\frac{S^{2}}{r}
\end{aligned}
\end{equation}
Equation Eq.~\ref{EQNR084} shows that the polar order surrounding the $+\frac{1}{2}-$defect is generated by the defect itself rather than being an independently imposed field. The $\lambda_{1}$ terms determine the additive polar forcing, while the $C$ and $K$ terms determine how the two polar harmonics relax and couple. In contrast, $\lambda_{2}$ does not directly generate the polar field and acts instead through the nematic equation.

\subsubsection{Defect Self Propulsion}

To determine the effect of the non-reciprocal terms on the $+\frac{1}{2}-$defect, we substitute Eq.~\ref{EQNR083} into Eq.~\ref{2101}, and use Eq.~\ref{EQNR081}. Again we only consider the non-reciprocal part to obtain
\begin{equation}\label{EQNR085}
\begin{aligned}
	R = &\lambda_{1}p_{0} \left[ \left( \partial_{r}S+\frac{S}{r} \right) + \left( \partial_{r}S-\frac{S}{r} \right)e^{2\mathrm{i}\phi} \right] + \left(2\lambda_{1}p_{1}\partial_{r}S\right) e^{\mathrm{i}\phi} - \left[ \lambda_{2} \left( (\partial_{r}S)^{2} - \frac{S^{2}}{r^{2}} \right) \right] e^{2\mathrm{i}\phi}
\end{aligned}
\end{equation}

The translational mode of the initial defect is again given by
\begin{equation}\label{EQNR086}
\begin{aligned}
	T_{x} = \partial_{x}f_{2}^{+\frac{1}{2}} = \frac{1}{2} \left[ \left( \partial_{r}S+\frac{S}{r} \right) + \left( \partial_{r}S-\frac{S}{r} \right)e^{2\mathrm{i}\phi} \right]
\end{aligned}
\end{equation}
Hence the first term of Eq.~\ref{EQNR085} is exactly
\begin{equation}
\begin{aligned}
	2\lambda_{1}p_{0}(r)T_{x}
\end{aligned}
\end{equation}
This is an important structural feature of the $\mathfrak{p}=2;\mathfrak{q}=1$ system. The defect-generated polar field couples directly onto the translational zero mode of the nematic $+\frac{1}{2}-$defect. This mirrors the advection that leads to defect self propulsion in active nematics, with the $f_1$ field playing the role of the velocity. 

The second set of terms in Eq.~\ref{EQNR085} are angularly orthogonal to $T_x$ and therefore do not contribute to translation along the defect polarity axis. The final set of terms generated by $\lambda_{1}$ and $\lambda_{2}$ can contribute. Using $\langle u,v\rangle \equiv \textrm{Re} \left[ \int d^{2}\mathbf{r}\, u^{*}v \right]$, the velocity is 
\begin{equation}\label{EQNR087}
\begin{aligned}
	V_{+\frac{1}{2}} &= - \frac{ \langle T_{x},R\rangle }{ \langle T_{x},T_{x}\rangle } \\
	&= - \frac{ 2\lambda_{1} \displaystyle \int d^{2}\mathbf{r}\  p_{0}(r)|T_x|^{2} }{ \langle T_{x},T_{x}\rangle } - \frac{ \pi \lambda_{2} \displaystyle \int_{0}^{\infty} rdr \ \left( \partial_{r}S-\frac{S}{r} \right) \left[  \left( (\partial_{r}S)^{2} - \frac{S^{2}}{r^{2}} \right) \right] }{ \langle T_{x},T_{x}\rangle }
\end{aligned}
\end{equation}
Unlike the $\mathfrak{p}=2;\mathfrak{q}=2$ Model, the defect velocity is therefore not, in general, a fixed linear combination of the $\lambda_{i}$ alone. It also depends on the secondary polar profiles $p_{0}$ and $p_{1}$ generated around the defect.
\newline
\newline
The direct $\lambda_{2}$ contribution can nevertheless be isolated. Since
\begin{equation}
\begin{aligned}
	(\partial_{r}S)^{2} - \frac{S^{2}}{r^{2}} = \left( \partial_{r}S+\frac{S}{r} \right) \left( \partial_{r}S-\frac{S}{r} \right)
\end{aligned}
\end{equation}
we obtain
\begin{equation}
\begin{aligned}
	V_{\lambda_{2}} = -C_{\lambda_{2}}\lambda_{2}
\end{aligned}
\end{equation}
where
\begin{equation}
\begin{aligned}
	C_{\lambda_{2}} = \frac{\pi}{ \langle T_{x},T_{x}\rangle } \int_{0}^{\infty} rdr \ \left( \partial_{r}S+\frac{S}{r} \right) \left( \partial_{r}S-\frac{S}{r} \right)^{2} >0
\end{aligned}
\end{equation}

Thus, for the phase convention used here, a positive $\lambda_{2}$ directly produces motion towards the head of the defect. The remaining velocity contributions are mediated through the secondary polar field.

Thus the general $\mathfrak{p}=2;\mathfrak{q}=1$ Model contains several possible mechanisms for defect motion and deformation. We now focus on the regime relevant to the leading non-reciprocal response, $\lambda_{1} >0$. $\lambda_{1}$ appears in both continuum equations. In the polar equation it contains the term
\begin{equation}
\begin{aligned}
	\frac{\lambda_{1}f_{0}}{2} \partial_{-}f_{2}
\end{aligned}
\end{equation}
which directly converts nematic gradients into polar order. In the nematic equation,
\begin{equation}
\begin{aligned}
	\lambda_{1} \left[ f_{1}\partial_{-}f_{2} + f_{-1}\partial_{+}f_{2} \right] = 2\lambda_{1} \left( P_{x}\partial_{x} + P_{y}\partial_{y} \right)f_{2}
\end{aligned}
\end{equation}
Thus, $\lambda_{1}$ both generates the secondary polar field and allows that polar field to advect the nematic texture.

\subsubsection{$\lambda_{1}-$Dominated Defect Self-Propulsion}
In the $\lambda_{1}-$dominated limit, the polar source Eq.~\ref{EQNR082} becomes
\begin{equation}\label{EQNR090}
\begin{aligned}
	\partial_{t} f_{1} \bigg|_{f_{1}=0} \simeq \frac{\lambda_{1}f_{0}}{2} \left( \partial_{r}S+\frac{S}{r} \right) - \lambda_{1} \frac{S^{2}}{r} e^{\mathrm{i}\phi}
\end{aligned}
\end{equation}
The leading polar contribution is the isotropic term,
\begin{equation}
\begin{aligned}
	\partial_{t} p_{0} \bigg|_{\textrm{source}} = \frac{\lambda_{1}f_{0}}{2} \left( \partial_{r}S+\frac{S}{r} \right)
\end{aligned}
\end{equation}
For the monotonic passive defect profile $ \left(\partial_{r}S+\frac{S}{r}\right)>0 $. Hence, 
\begin{equation}
\begin{aligned}
	\lambda_{1} > 0 \quad \Rightarrow \quad p_{0}>0
\end{aligned}
\end{equation}
Since a positive real $f_{1}$ points along the positive $x-$direction, and the positive $x-$direction is the tail direction for our defect convention, the $+\frac{1}{2}-$defect generates a polar field pointing towards its tail.
\newline
\newline
The $\lambda_{1}$ term in the non-reciprocal parts of Eq.~\ref{eq:f2_21} can be written
\begin{equation}\label{EQNR091}
\begin{aligned}
	R_{2}^{(\lambda_{1})} = 2\lambda_{1}p_{0}(r) \cdot T_{x} + 2\lambda_{1}p_{1}(r) \cdot \partial_{r}S e^{\mathrm{i}\phi}
\end{aligned}
\end{equation}
The second term is angularly orthogonal to $T_{x}$. Consequently,
\begin{equation}\label{EQNR092}
\begin{aligned}
	v_{+\frac{1}{2}} = - 2\lambda_{1} \frac{ \displaystyle \int d^{2}\mathbf{r} \  p_{0}(r)|T_{x}|^{2} }{ \langle T_{x},T_{x}\rangle }
\end{aligned}
\end{equation}
It is useful to define the translational-mode weighted polar field
\begin{equation}
\begin{aligned}
	\bar{p}_{0} = \frac{ \displaystyle \int d^{2}\mathbf{r}\  p_{0}(r)|T_{x}|^{2} }{ \langle T_{x},T_{x}\rangle }
\end{aligned}
\end{equation}
so that
\begin{equation}\label{EQNR093}
\begin{aligned}
	v_{+\frac{1}{2}} = -2\lambda_{1}\bar{p}_{0}
\end{aligned}
\end{equation}
For $\lambda_{1}>0$, the defect-generated polar field has $\overline{p}_{0}>0$. Therefore,
\begin{equation}
\begin{aligned}
	v_{+\frac{1}{2}}<0
\end{aligned}
\end{equation}
and hence
\begin{equation}
\begin{aligned}
	\lambda_{1}>0 \quad\Rightarrow\quad \textrm{$+\frac{1}{2}-$defect moves towards its head}
\end{aligned}
\end{equation}
The physical meaning of Eq.~\ref{EQNR093} can also be seen by writing
\begin{equation}
\begin{aligned}
	\partial_{t}f_{2} \bigg|_{\lambda_{1}} = 2\lambda_{1} \mathbf{P}\cdot\nabla f_{2}
\end{aligned}
\end{equation}
In the usual advection form,
\begin{equation}
\begin{aligned}
	\partial_{t}f_{2} + \mathbf{u}\cdot\nabla f_{2} = \cdot \cdot \cdot 
\end{aligned}
\end{equation}
the corresponding local velocity is
\begin{equation}
\begin{aligned}
	\mathbf{u} = -2\lambda_{1}\mathbf{P}
\end{aligned}
\end{equation}
Thus, for $\lambda_{1}>0$, a tail-directed polar field advects the nematic texture in the opposite direction, towards the defect head.

Equation Eq.~\ref{EQNR093} is the exact leading dependence of the defect velocity on $\lambda_{1}$ for a given secondary polar field. Since $f_{1}$ is itself a secondary slow field, $\overline p_{0}$ should in general be determined from its own dynamics rather than algebraically eliminated. At fixed polar profile the defect velocity is linear in $\lambda_{1}$. If instead the system is initialized at exactly $f_{1}=0$ and the polar field is treated perturbatively at very short times, then $p_{0}\propto\lambda_{1}$ and the earliest feedback of the induced polar field onto the nematic defect begins at order $\lambda_{1}^{2}$.
\subsubsection{Defect Shape for $\lambda_{1}>0$}
We will now determine how the same $\lambda_{1}-$driven mechanism changes the internal shape of the $+\frac{1}{2}-$defect. The central distinction is between the mean polar advection, which translates the defect, and the spatial variation of this advection, which changes its internal director texture.

Using Eq.~\ref{EQNR091} and Eq.~\ref{EQNR093}, the translation-subtracted residual is
\begin{equation}\label{EQNR094}
\begin{aligned}
	R_{\textrm{shape}} & = R_{2}^{(\lambda_{1})} + V_{+\frac{1}{2}}T_{x} \\
	&= 2\lambda_{1} \left[ p_{0}(r)-\overline{p}_{0} \right]T_{x} + 2\lambda_{1}p_{1}(r) \partial_{r}S e^{\mathrm{i}\phi}
\end{aligned}
\end{equation}
Using
\begin{equation}
\begin{aligned}
	e^{-\mathrm{i}\phi}T_{x} = \partial_{r}S\cos(\phi) - \mathrm{i} \frac{S}{r}\sin(\phi)
\end{aligned}
\end{equation}
we obtain
\begin{equation}\label{EQNR095}
\begin{aligned}
	e^{-\mathrm{i}\phi} R_{\textrm{shape}} &= 2\lambda_{1}p_{1} \cdot \partial_{r}S + 2\lambda_{1} \left[ p_{0}-\bar{p}_{0} \right]  \cdot \partial_{r}S\cos(\phi) - 2\mathrm{i}\lambda_{1} \left[ p_{0}-\bar{p}_{0} \right] \cdot  \frac{S}{r}\sin(\phi)
\end{aligned}
\end{equation}
Therefore,
\begin{equation}\label{EQNR096}
\begin{aligned}
	\partial_tS \bigg|_{\textrm{shape}} &= 2\lambda_{1}p_{1} \cdot \partial_{r}S + 2\lambda_{1} \left[ p_{0}-\bar{p}_{0} \right] \cdot \partial_{r}S\cos(\phi) \\
	\partial_t\theta \bigg|_{\textrm{shape}} &= - \frac{\lambda_{1}}{r} \left[ p_{0}(r)-\bar{p}_{0} \right]\cdot  \sin(\phi)
\end{aligned}
\end{equation}

As in the $\{2,2\}$ case, we concentrate now on the orientational changes given by the second equation above. This once again dominated by the $p_0$ contributions. As before we write for small times $t$,
\begin{equation}\label{EQNR097}
\begin{aligned}
	\theta(r,\phi) = \frac{\phi}{2} + t\delta(r)\sin(\phi).
\end{aligned}
\end{equation}
Where 
\begin{equation}
\begin{aligned}
	\delta(r) = \frac{\lambda_{1}}{r} \left[ \bar{p}_{0}-{p}_{0}(r) \right].
\end{aligned}
\end{equation}

We assume $t$ is small and the balance between bend and splay is again given by Eq.~\ref{EQNR074} leading to the equation
\begin{equation}
\delta\phi = 2t\left[{\delta(r)}+2r\partial_r\delta(r)\right].
\end{equation}

Hence,
\begin{equation}
\begin{aligned}
	\delta+2r\partial_{r}\delta<0 &\quad\Rightarrow\quad \textrm{bend-dominated head expands} \\
	\delta+2r\delta_{r}\epsilon>0 &\quad\Rightarrow\quad \textrm{splay-dominated tail expands}
\end{aligned}
\end{equation}

In the regime in which $\lambda_1>0$ we expect the splay dominated region to expand, exactly as is shown in our simulations. 

The physical origin of this deformation is the spatially non-uniform advection generated by the defect-induced polar field. If $p_{0}$ were spatially uniform, the $\lambda_{1}$ term would produce only rigid translation. Since the defect generates a polar field that is strongest close to the defect and decreases outward, the nematic texture is advected by different amounts at different radii. After the rigid translational component is removed, this spatial variation remains as an internal orientational deformation.
\newline
\newline
Thus, in the $\lambda_{1}-$dominated regime,
\begin{equation}
\begin{aligned}
	\lambda_{1}>0 \quad\Rightarrow\quad \begin{cases}
		\textrm{$+\frac{1}{2}-$defect generates polar order directed towards its tail}\\
		\textrm{$+\frac{1}{2}-$defect moves towards its head}\\
		\textrm{splay-dominated tail region expands}\\
	\end{cases}
\end{aligned}
\end{equation}

\end{document}